\documentclass[a4paper,twocolumn,11pt]{quantumarticle}
\pdfoutput=1
\usepackage[utf8]{inputenc}
\usepackage[english]{babel}
\usepackage[T1]{fontenc}
\usepackage{amsmath}
\usepackage{hyperref}

\usepackage{amsmath}
\usepackage{amsthm}
\usepackage{physics}
\usepackage{amsfonts}
\usepackage{enumitem}
\usepackage{amssymb}
\usepackage{bbm}
\usepackage{float}
\usepackage{cprotect}
\usepackage{mathtools}
\usepackage[toc]{appendix}
\usepackage[linesnumbered,ruled,vlined]{algorithm2e}

\newtheorem{theorem}{Theorem}[section]

\newtheorem{lemma}{Lemma}[section]

\newtheorem{example}{Example}[section]
\newtheorem{definition}{Definition}[section]

\DeclareMathOperator*{\argmax}{argmax}
\DeclareMathOperator*{\argmin}{argmin}

\newcommand{\ketx}{\ket{\u{x}}} 
\newcommand{\brax}{\bra{\u{x}}}

\newcommand{\ux}{\underline{x}}

\renewcommand{\u}[1]{\underline{#1}}

\newcommand{\maxb}{\max\limits_{\u{x} \in \{0, 1 \}^n}}
\newcommand{\sumb}{\sum\limits_{\u{x} \in \{0, 1 \}^n}}

\newcommand{\bn}{\{0, 1\}}

\newcommand{\bd}{\begin{definition}}
\newcommand{\ed}{\end{definition}}
\newcommand{\bex}{\begin{example}}
\newcommand{\eex}{\end{example}}
\newcommand{\bt}{\begin{theorem}}
\newcommand{\et}{\end{theorem}}
\newcommand{\bl}{\begin{lemma}}
\newcommand{\el}{\end{lemma}}
\newcommand{\ham}{\hat{\mathcal{H}}}
\newcommand{\un}{\hat{\mathcal{U}}}
\newcommand{\unc}{\un_{\mathcal{C}}}

\newcommand{\uncsn}{\un_{\mathcal{C}^{\text{NAE}}_{\us}}}
\newcommand{\csn}[1]{\mathcal{C}^{\text{NAE}}_{\us}(#1)}
\newcommand{\unb}{\un_{\mathcal{B}}}
\newcommand{\hamc}{\ham_{\mathcal{C}}}

\newcommand{\hamcsn}{\ham_{\mathcal{C}^{\text{NAE}}_{\us}}}
\newcommand{\hamb}{\ham_{\mathcal{B}}}
\newcommand{\us}{\underline{\sigma}}

\begin{document}

\title{Quantum Approximate Optimisation for Not-All-Equal SAT}

\author{Andrew Elkadi}
\affiliation{Department of Computing, Imperial College London, London SW7 2AZ, United Kingdom.}
\author{Roberto Bondesan}
\affiliation{Department of Computing, Imperial College London, London SW7 2AZ, United Kingdom.}
\maketitle

\begin{abstract}
Establishing quantum advantage for variational quantum algorithms is an important direction in quantum computing. In this work, we apply the Quantum Approximate Optimisation Algorithm (QAOA) 
-- a popular variational quantum algorithm for general combinatorial optimisation problems -- 
to a variant of the satisfiability problem (SAT): Not-All-Equal SAT (NAE-SAT). We focus on regimes where the problems are known to have solutions with low probability and introduce a novel classical solver that outperforms existing solvers. Extensively benchmarking QAOA against this, we show that while the runtime of both solvers scales exponentially with the problem size, the scaling exponent for QAOA is smaller for large enough circuit depths. This implies a polynomial quantum speedup for solving NAE-SAT. 
\end{abstract}

\section{Introduction}

Quantum algorithms provide a provable advantage for a handful of application areas, including the simulation of quantum physics and cryptography \cite{dalzell2023quantum}. A central direction in the field is to find new tasks that can be sped up with a quantum computer.
The current NISQ era of quantum devices has motivated the investigation of variational quantum algorithms, which are expected to be more resilient to noise \cite{Cerezo_2021}.
Variational quantum algorithms for tasks such as machine learning, quantum chemistry simulation and combinatorial optimisation have been proposed \cite{Cerezo_2021}.
One of the most prominent examples is the Quantum Approximate Optimisation Algorithm (QAOA) \cite{farhi14,shaydulin19}. QAOA is designed to find approximate solutions to general combinatorial optimisation problems, based on a Trotterised adiabatic process.

However, variational quantum algorithms are heuristic in nature and without a large enough quantum device to test them, it is unclear whether they can actually provide any practical advantage.
In fact, it can be shown that optimising the variational parameters is NP-hard in the worst case \cite{Bittel_2021}.
Further, recent work has analytically studied the approximation ratio of QAOA for spin glasses and related random combinatorial problems in the limit of a large number of qubits at fixed circuit depths 
\cite{farhi2020quantum,Farhi2022quantumapproximate,https://doi.org/10.48550/arxiv.2204.10306,chou2022limitations}. This work has found that the performance of QAOA is away from optimality. 
A more positive set of results were instead obtained for the
Boolean satisfiability problem (SAT) \cite{boulebnane22}.

SAT is the canonical NP-complete problem \cite{stephen71} with a wide range of applications that include circuit design \cite{hong10}, logic-based planning \cite{kautz92} and bug detection \cite{jackson00}.
While it is unlikely that a polynomial-time solving algorithm will ever be found, there is interest in finding \textit{efficient enough} algorithms \cite{zhang02}.
An instance of SAT is a Boolean formula consisting of $n$ variables arranged in $m$ clauses. Each clause represents a constraint on the variables, and to solve SAT is to determine whether there exists a variable assignment that satisfies all the constraints. When each clause is restricted to $k$ variables the resulting problem is known as $k$-SAT \cite{moore11}.

Of particular interest are instances with \textit{clause densities} $r = m/n$ that undergo \textit{computational phase transitions} \cite{moore11}. The \textit{satisfiability ratio} $r_k$ describes a critical value of $r$ beyond which instances are unlikely to have any solutions. On the other hand, the \textit{algorithmic ratio} $a_k$ is a threshold beyond which no classical algorithm is known to find solutions efficiently. Notably, it is known that the algorithmic ratio is strictly smaller than the satisfiability ratio ($a_k < r_k$) \cite{coja09}.
It is thus natural to look for quantum advantage in the regime $a_k < r \leq r_k$. 

The recent work of Boulebnane \& Montanaro \cite{boulebnane22} explores the \textit{success probability} and \textit{median running time} of QAOA on $k$-SAT instances at $r_k$, discovering empirically a quantum advantage over classical algorithms. 
They benchmark QAOA against a range of classical solvers and find that it outperforms the best performing algorithm, \verb|WalkSATlm| \cite{cai14}. 
Previous works on QAOA for SAT \cite{Akshay_2020, zhang2022quantum} connect 
phase transitions to Barren plateaus -- regions where parameter training is infeasible -- and identify a quantum advantage in the approximation ratio for MAX-$1$-$k$-SAT$^+$ compared to a classical greedy baseline for this problem. This is however for $10$ variables and large circuit depths, meaning it is unclear how these results will scale.

In this work, we extend the numerical analysis of \cite{boulebnane22} to the Not-All-Equal SAT (NAE-SAT) problem, providing a new example where QAOA empirically outperforms the best known classical solvers for a random constraint satisfaction problem.
NAE-SAT is a variant of SAT where we additionally restrict the problem, such that no clause contains variables that are all satisfied.
NAE-SAT is also NP-complete and
plays an important role in proving reductions in computational complexity \cite{moret1988planar}. Special cases of NAE-SAT can be interpreted as classical computer science problems, such as bicolorability testing of hypergraphs where
there exists no monochromatic hyperedge, or as the set splitting problem.
The phase diagram for random $k$-NAE-SAT
has been extensively studied \cite{ding2013satisfiability,sly16}.
We make the following contributions: 
\begin{itemize}[noitemsep, nolistsep]
    \item Introduce \verb|WalkSATm2b2|, a new solver for $k$-NAE-SAT, that extends on \verb|WalkSATlm| and accounts for the symmetry of the problem's instances. We empirically show that it outperforms \verb|WalkSATlm| on $k$-NAE-SAT.
    \item Demonstrate that QAOA outperforms both \verb|WalkSATlm| and \verb|WalkSATm2b2| on a range of $k$-NAE-SAT instances. 
    \item Produce novel \textit{success probability} and \textit{median running time} results of QAOA for a range of $k$-NAE-SAT problems. We also consider the scaling of these heuristics and interpret their \textit{excessive scaling} performance.
    \item Develop a library allowing for the efficient verification of $k$-NAE-SAT results and simulation of other random problems, including $k$-SAT.
\end{itemize}

\section{Background}
\subsection{Quantum Approximate Optimisation Algorithm}
The Quantum Approximate Optimisation Algorithm (QAOA) is a Variational Quantum Algorithm whose origins lie in Adiabatic Quantum Computing \cite{farhi00}. To solve an optimisation problem, QAOA \textit{represents} the objective function $\mathcal{C}: \bn^n \rightarrow \mathbb{R}$, with a \textit{problem} Hamiltonian $\hamc$ such that $\hamc \ketx = \mathcal{C}(\ux)\ketx$. As such, finding $\max_{\ux \in \bn^n} \mathcal{C}(\ux)$ corresponds to finding $\max_{\ux \in \bn^n} \expval{\ham_{\mathcal{C}}}{\ux}$, the maximal eigenstate of $\hamc$. QAOA guides the search for such a state through the alternating application of two parameterised operators
\begin{equation}
\begin{split}
    \unb(\beta_j) &= e^{-i\beta_j \hamb}\\
    \unc(\gamma_j) &= e^{-i\gamma_j \hamc}
\end{split}
\end{equation}
where $\hamc$ is as before, and $\hamb$ is the Hamiltonian with highest energy state $\ket{+}^{\otimes n} \equiv \frac{1}{\sqrt{2^n}} \sum_{\ux \in \bn^n} \ketx$. The output of the QAOA circuit gives 
\begin{equation}
    \ket{\u{\gamma}, \u{\beta}}_p := \prod\limits_{i = 0}^{p-1}\unb(\beta_i)\unc(\gamma_i)\ket{+}^{\otimes n}
\end{equation}  
where $p$ is known as the circuit's \textit{depth}. The parameters $\u{\beta}, \u{\gamma}$ are selected to maximise $F_p := \bra{\u{\gamma}, \u{\beta}}\hamc\ket{\u{\gamma}, \u{\beta}}_p$, where adiabatic considerations guarantee \cite{farhi14}:
\begin{equation}
    \lim\limits_{p \rightarrow \infty} \max\limits_{\u{\gamma},\u{\beta}} F_p = \maxb C(\ux)
\end{equation}

\subsection{$k$-NAE-SAT}
The Boolean satisfiability problem (SAT) seeks a \textit{satisfying assignment} for a Boolean propositional formula $\phi$ in Conjunctive Normal Form (CNF) \cite{stephen71}. Given $n$ Boolean literals (a Boolean variable $x \in \bn$ or its negation $\neg x$), a formula in CNF consists of a disjunction of $m$ clauses $\phi = \bigwedge_{i=0}^{m-1} c_i$, where each clause $c_i$ is a conjunction of $k_i$ of the Boolean literals $c_i = \bigvee_{j = 0}^{k_i - 1} l_{ij}$. As such, a clause is satisfied (takes the value of $1$) if any one of the literals within is satisfied, while a CNF formula is satisfied if all the clauses within are satisfied. When all clauses contain exactly $k$ literals ($\forall i, k_i = k$), the problem is known as $k$-SAT. Additionally constraining the problem such that no clause can contain literals that are all satisfied gives $k$-NAE-SAT.
We note that NAE-SAT has a $\mathbb{Z}_2$ symmetry under global bit flip.
If $c_i$ is satisfied but not all $l_{ij}$ are true, we can flip all $l_{ij}$ and still have $c_i$ satisfied, since those $l_{ij}$'s that were false are now true and vice versa.

\subsection{Computational Phase Transitions}
In this work, we consider solving randomly generated $k$-NAE-SAT problems and are interested in two of their quantities: the \textit{satisfiability ratio} and \textit{algorithmic ratio}. To construct a random CNF formula of $n$ variables, $m$ clauses and $k$ variables per clause, for each clause, with replacement, $k$ variables are chosen with uniform randomness from $n \choose k$ combinations. Each variable is randomly negated with probability $p = 0.5$ \cite{moore11}. An instance, denoted $F_k(n, m)$, is said to be satisfiable/unsatisfiable \textit{with high probability} (w.h.p) if $\lim_{n \rightarrow \infty} P[F_k(n, m) \text{ satisfiable/unsatisfiable}] = 1$. 

The clause densities, $r = m/n$, of these instances allow us to define the quantities of interest. In particular, the satisfiability ratio, $r_k(n)$, is a sharp threshold such that $F_k(n, m)$ instances are satisfiable w.h.p $\forall \varepsilon > 0: r < r_k(n) - \varepsilon$, yet unsatisfiable w.h.p $\forall \varepsilon > 0: r > r_k(n) + \varepsilon$. For $k$-NAE-SAT, $r_k(n)$ is not known precisely for $k \geq 3$. Yet, it has been shown that $2^{k-1}\ln(2) - \order{1} \leq r_k(n) \leq 2^{k-1}\ln(2)$, and, while $r_k(n)$ is dependent on $n$, it is conjectured to converge in the limit for fixed $k$ \cite{coja09}.

Besides the satisfiability ratio, we are interested in problems that are satisfiable but cannot be solved \textit{efficiently} with currently known classical algorithms. In particular, the algorithmic ratio $a_k$ is a sharp threshold such that there are no efficient (polynomial time) known classical solvers for $F_k(n, m)$ instances with $r > a_k$. For $k$-NAE-SAT $a_k = 2^{k-1}\ln(k)/k$ \cite{coja09}.

\subsection{WalkSAT and WalkSATlm}
Classical algorithms for solving SAT can be split into two main categories: Complete algorithms and Stochastic Local Search (SLS) algorithms \cite{cai14}. SLS algorithms are \textit{incomplete} in the sense that they cannot determine with certainty whether a given formula is satisfiable. However, in the regime of necessarily satisfiable problems, they are very efficient, particularly for randomly generated instances. Their general approach consists of starting with a random assignment and flipping variable truth values until a satisfying assignment is found. Deciding which variables to flip is done through a combination of randomness or \textit{locally searching} the space of truth assignments to optimise a constructed heuristic.

In general, SLS algorithms select variables to flip based on optimising \textit{scoring functions}. A \textit{focused random walk} is an SLS approach that chooses variables only from clauses that are unsatisfied. \verb|WalkSAT| \cite{selman99}, is a popular example that employs a variable selection scheme based on $break$ scores (Algorithm \ref{walksat}). For a variable $x$ in a Boolean formula $\phi$, $break(x)$ is the number of clauses that become unsatisfied by flipping $x$, while $make(x)$ is defined as the number of clauses that become satisfied by flipping $x$. At each iteration, any \textit{freebie} variables are flipped (variables with \textit{break} scores of 0). Otherwise, a Bernoulli random variable (controlled by \textit{noise} parameter $p$) is sampled. Depending on the outcome, a variable to be flipped is either chosen at random or amongst those that have lowest (non-zero) $break$ scores in the clause. \verb|WalkSAT| solves \textit{tiebreaks} (multiple variables with optimal score) through random selection. We note the inclusion of a \textit{max\_flips} argument to account for SLS being incomplete, avoiding the algorithm looping infinitely in the case of an unsatisfiable formula.

\verb|WalkSATlm| \cite{cai14}, builds on \verb|WalkSAT|, altering the tiebreak procedure to no longer be random. A new scoring function is introduced as follows.

\bd ($\tau$-satisfied) Given a Boolean formula $\phi$ and assignment $\alpha$, a clause $c$ in $\phi$ is $\tau$-satisfied iff under $\alpha$, it contains exactly $\tau$ satisfied literals \cite{cai14}. 
\ed

As such, a $0$-satisfied clause is an unsatisfied clause, while $\geq1$-satisfied clauses are satisfied. $1$-satisfied clauses are of interest as they are very unstable - flipping the variable corresponding to the satisfied literal returns the clause to an unsatisfied state. Using this, the $make$ score is generalised.

\bd ($\tau$-level make) For a variable $x$ in a Boolean formula $\phi$, $make_\tau(x)$ is the number of $(\tau - 1)$-satisfied clauses that become $\tau$-satisfied by flipping $x$ \cite{cai14}.
\ed

It is clear that $make_1$ is equivalent to $make$. $make_2$, whereas, is the number of clauses that move away from instability - flipping $x$ would make them 2-satisfied and so they can no longer be broken by flipping exactly one variable. This corresponds to the number of variables that would have their $break$ scores decreased by flipping $x$. \verb|WalkSATlm| introduces the $lmake$ scoring function to be used in tiebreak situations
\begin{equation}
    lmake(x) = \omega_1 \cdot make_1(x) + \omega_2 \cdot make_2(x)
\end{equation}
where $\omega_1, \omega_2$ are hyperparameters (Algorithm \ref{walksatlm}). The variable with the maximal $lmake$ score amongst those with the minimal $break$ scores is chosen to be flipped. This simple change allows \verb|WalkSATlm| to outperform \verb|WalkSAT| by orders of magnitude for $k > 3$ and stems from the fact that tiebreaks occur in $40\%$ and $30\%$ of steps in $5$-SAT and $7$-SAT respectively \cite{cai14}.  
To the best of our knowledge, no SLS algorithm adapted to NAE-SAT has been discussed in the literature. We will show below how to modify \verb|WalkSATlm| to boost its performance for NAE-SAT.

\begin{algorithm}[h] 
  \SetKwFunction{WalkSAT}{WalkSAT}
  \SetKwProg{Fn}{Function}{:}{}

  \Fn{\WalkSAT{$\phi$, $p$, $max\_flips$}}{
    Randomly assign truth values to all variables in $\phi$\;
    \For{$i$ = 1 to $max\_flips$}{
      \If{$\phi$ is satisfied}{
        \Return{the assignment}\;
      }
      Choose uniformly at random an unsatisfied clause $c$ from $\phi$\;
      \If{$\exists x \in c: break(x) = 0$}{
        Flip the value of (the first such) $x$\tcp*{'Freebie'}
      }
      Sample $X \sim Bernoulli(p)$\;
      \If{X}{
        Choose uniformly at random $x \in c$ and flip its value\;
      }
      \Else{
        \textbf{let} $B = \argmin_{y \in c} \text{break}(y)$\;
        Choose uniformly at random $x \in B$ and flip its value\;
      }
    }
    \Return{No satisfying assignment found}\;
  }

  \caption{WalkSAT Algorithm \label{walksat}}
\end{algorithm}
\begin{algorithm}[h] 
  \SetKwFunction{WalkSATlm}{WalkSATlm}
  \SetKwProg{Fn}{Function}{:}{}

  \Fn{\WalkSATlm{$\phi$, $p$, $max\_flips$}}{
    Randomly assign truth values to all variables in $\phi$\;
    \For{$i$ = 1 to $max\_flips$}{
    \tcp{As in WalkSAT}
      Sample $X \sim Bernoulli(p)$\;
      \If{X}{
        Choose uniformly at random $x \in c$ and flip its value\;
      }
      \Else{
        \textbf{let} $B = \argmin_{y \in c} \text{break}(y)$\;
        \textbf{let} $M = \argmax_{v \in B} \text{lmake}(v)$\;
        Choose uniformly at random $x \in M$ and flip its value\;
      }
    }
    \Return{No satisfying assignment found}\;
  }

  \caption{WalkSATlm Algorithm \label{walksatlm}}
\end{algorithm}

\subsection{QAOA for $k$-SAT}
The recent work of Boulebnane \& Montanaro \cite{boulebnane22} explores the use of \textit{fixed angle, constant depth} QAOA for solving $k$-SAT. While QAOA is often used for finding approximate solutions, this work examines its application as an exact solver. Namely, the authors consider repeatedly running QAOA until a measurement outcome on the output state corresponds to a satisfying assignment. This translates into an exact, yet incomplete, algorithm for solving SAT. If no satisfying assignment exists then no measurement outcome will be a solution and the algorithm will infinitely loop. Boulebnane \& Montanaro prove the existence of an analytic formula for the probability of success of the algorithm and empirically demonstrate that this is closely related to its median running time. They further benchmark QAOA against a number of classical $k$-SAT solvers and identify a quantum advantage.
We will review below the definitions of probability of success and median running time in our context.

\section{Methods}
\subsection{Random Problems}
We extend the procedure of random problem generation from the work of \cite{boulebnane22}.
\bd \label{ksatgen}
(Random $k$-NAE-SAT generation). Let $k \in \mathbb{N}^+$ and $r > 0$. To form a random $k$-NAE-SAT $\us \sim CNF(n, k, r)$ instance:
\begin{itemize}
    \item Sample $m \sim \text{Poisson}(rn)$
    \item Generate $\us := \bigwedge_{i = 0}^{m-1} \sigma_i$, $\us \sim F_k(n, m)$
\end{itemize}
\ed

\subsection{Problem Hamiltonian}
Finding a satisfying $k$-NAE-SAT assignment corresponds to maximising the number of satisfied clauses, or equivalently, minimising the number of unsatisfied clauses. Solving for problem instances $\underline{\sigma} \sim CNF(n, k, r)$, with $m$ clauses, such that 
\begin{equation}
    \underline{\sigma} := \bigwedge\limits_{i = 0}^{m-1} \sigma_i
    \,,\quad 
    \sigma_i = \bigvee_{j = 0}^{k-1} l_{\sigma_{ij}}
\end{equation}
we denote an assignment $\ux$ satisfying $\us$ with the NAE constraint by $\ux \vdash_{\text{NAE}} \us $.
Our objective function to be minimised is therefore
\begin{equation}
\mathcal{C}^{\text{NAE}}_{\us}(\ux) = \sum\limits_{i = 0}^{m-1} \mathbbm{1}\{\ux \nvdash_{\text{NAE}} \sigma_i\}
\end{equation}
This is represented by the diagonal operator
\begin{equation}
    \hamcsn = \sumb \mathcal{C}^{\text{NAE}}_{\us}(\ux) \ketx\brax
\end{equation}
with parameterised unitary operator
\begin{equation}
    \uncsn(\gamma) = e^{-i\gamma\hamcsn} = \sumb e^{-i\gamma \mathcal{C}^{\text{NAE}}_{\us}(\ux)} \ket{\ux}\bra{\ux}
\end{equation}

The gate based implementation of this operator can be found in Appendix  \ref{qaoagatebased}. 

\subsection{QAOA Procedure}
We denote the output of the QAOA circuit by
\begin{equation}
\ket{\Psi^{\text{NAE}}(\u{\beta}, \u{\gamma}, \us)}_p := \prod\limits_{i = 0}^{p-1}\unb(\beta_i)\uncsn(\gamma_i)\ket{+}^{\otimes n}
\end{equation}
This is measured and the corresponding outcome is checked as a satisfiable assignment. We repeat this procedure until a satisfying assignment is found (if one exists). 

We evaluate the effectiveness of this procedure for \textit{fixed angle} QAOA on instances with clause densities near the satisfiability threshold $r_k$. Asymptotic values of the threshold are known to be 
\begin{equation} \label{satthresnae}
    r^{\text{NAE}}_{k} = \left( 2^{k-1} - \frac{1}{2} - \frac{1}{4\log 2}\right)\log 2 + o_k(1)
\end{equation}
where $o_k(1) \rightarrow 0$ in the $k \rightarrow \infty$ limit \cite{nam21}. As such, we approximate 
\begin{equation}
    \hat{r}^{\text{NAE}}_{k} = \left( 2^{k-1} - \frac{1}{2} - \frac{1}{4\log 2}\right)\log 2
\end{equation}

Next, we introduce the notion of the \textit{success probability}.

\bd ($k$-NAE-SAT success probability) Let $k, n \in \mathbb{N}^+, r > 0$ in $ \underline{\sigma} \sim CNF(n, k, r)$ and $\underline{\beta}, \underline{\gamma} \in \mathbb{R}^p$, the success probability of a random $k$-NAE-SAT QAOA instance is defined as
\begin{equation} \label{psuccnae}
\begin{split}    
    &p^{\text{NAE}}_{succ}(\u{\beta}, \u{\gamma}, \us) = \\
    &{}_p
    \bra{
    \Psi^{\text{NAE}}(\underline{\beta},\underline{\gamma}, \us)
    }
    \{ \hamcsn = 0 \}
    \ket{\Psi^{\text{NAE}}(\underline{\beta},\underline{\gamma}, \us)}_p
\end{split}
\end{equation}
\ed
where $\{ \hamcsn = 0\}$ is the projector onto the subspace of states with eigenvalues 0 (satisfying states). For brevity, this is denoted as $p^{\text{NAE}}_{succ}(\us)$.

The fixed angles are pretrained parameters $\underline{\beta}^*$ and $\underline{\gamma}^*$. For each layer count $p$, and value of $k$, they are selected by:
\begin{enumerate}
    \item Generating $t = 100$ random instances,\\ $\{ \us_i : \us_i \sim CNF(n=12, k, \hat{r}^{\text{NAE}}_k)\}_{i = 0}^{t - 1}$
    \item Initialising $\beta_i = 0.01, \gamma_i = -0.01$
    \item Optimising $\underline{\beta}, \underline{\gamma}$ for 100 epochs using the \verb|PyTorch| ADAM optimiser to maximise the success probability over the instances
    \begin{equation}
        \frac{1}{t}\sum_{i = 0}^{t-1} p^{\text{NAE}}_{succ}(\us_i)
    \end{equation}
\end{enumerate}

We emphasise that the parameters are selected through training only on $CNF(\vb{n=12},k,\hat{r}^{\text{NAE}}_k)$ instances. In doing so, we assess QAOA's ability to generalise across instances with other variable counts $n$.

To evaluate the circuits, $2500$ \textbf{satisfiable} random $k$-NAE-SAT instances, $\{ \us_i : \us_i \sim CNF(n, k, \hat{r}^{\text{NAE}}_k)\}$ are generated. Importantly, this is done without biasing the procedure (e.g. by selecting variables in such a way that the final formula is guaranteed to be satisfiable). Instead, we randomly generate the instances then confirm if they are satisfiable using the \verb|Glucose4| solver from \verb|PySAT| \cite{pysatcode}. This is an efficient, complete $k$-SAT solver. As such, to use \verb|Glucose4|, we recast $k$-NAE-SAT in terms of $k$-SAT and determine with certainty if the instance is satisfiable (see Appendix \ref{naesatassat}).

\subsection{Efficient Classical Simulation}
The operator $\uncsn$ can be applied in $\order{2^k k N\log{N}}$ operations, where $N=2^n$ (see Appendix \ref{qaoacost}). Amortising the cost of calculating $\mathcal{C}^{\text{NAE}}_{\us}(\ux)$ through preprocessing allows for applications of the operator in $\order{N}$ operations.

The mixing unitary operator
\begin{equation}
\unb(\beta) = e^{i\beta \hamb} = e^{i\beta \sum\limits_{j = 0}^{n-1} X_j }
\end{equation}
can be applied in $\order{N \log N}$ operations, where $N=2^n$ (see Appendix \ref{qaoacost}). 

\subsection{Classical Benchmarking}
To account for the additional symmetry in $k$-NAE-SAT, we adapt the \verb|WalkSATlm| algorithm and introduce \verb|WalkSATm2b2|. \verb|WalkSATlm| uses the $lmake$ score during tiebreaks, as such, it only considers the impact flipping a variable has on increasing the number of satisfied literals. This approach is relevant for $k$-SAT, where formulas are more likely to be satisfied if more literals are set true. On the other hand, $k$-NAE-SAT requires that at least one literal be unsatisfied in each clause. To take this into account, for a variable $x$ in a Boolean formula $\phi$ we introduce $make^{NAE}(x)$ as the number of clauses that become satisfied by flipping $x$, while $break^{NAE}(x)$ as the number of clauses that become unsatisfied by flipping $x$, in the NAE formulation. Further, we introduce the notion of a $\tau$-level break.

\bd ($\tau$-level break) For a variable $x$ in a Boolean formula $\phi$, $break_\tau(x)$ is the number of $(\tau)$-satisfied clauses that become $(\tau - 1)$-satisfied by flipping $x$.
\ed

Consequently, we deduce the following relationships 
\begin{equation}
\begin{split} \label{naerelat}
    make^{NAE}(x) = make_1(x) + break_k(x) \\
    break^{NAE}(x) = break_1(x) + make_k(x)
\end{split}
\end{equation}
These follow from the fact that in $k$-NAE-SAT, satisfying all the literals now means the clause is no longer satisfied. It is clear that unlike $k$-SAT, where $make_1(x) = make(x)$, $make_1(x) \leq make^{NAE}(x)$.

We construct a new scoring function
\begin{equation}
\begin{split}
    m2b2(x) &= \omega_1 \cdot make^{NAE}(x) \\ + &\omega_2 \cdot (make_2(x) + break_{k-1}(x)) \\
    &= \omega_1 \cdot (make_1(x) + break_k(x)) \\ + &\omega_2 \cdot (make_2(x) + break_{k-1}(x))
\end{split}
\end{equation}
where $\omega_1, \omega_2$ are hyperparameters.

This form allows for an immediately efficient implementation of the scoring function where the algorithm keeps track of the number of true literals in each clause and considers changes caused by a variable being flipped. The \verb|WalkSATm2b2| algorithm uses $m2b2$ to resolve tiebreaks, regaining symmetry in its selection procedure.

\begin{algorithm}[h]
  \SetKwFunction{WalkSATmb}{WalkSATm2b2}
  \SetKwProg{Fn}{Function}{:}{}

  \Fn{\WalkSATmb{$\phi$, $p$, $max\_flips$}}{
    Randomly assign truth values to all variables in $\phi$\;
    \For{$i$ = 1 to $max\_flips$}{
    \tcp{As in WalkSATlm}
      Sample $X \sim Bernoulli(p)$\;
      \If{X}{
        Choose $x \in c$ randomly and flip its value\;
      }
      \Else{
        \textbf{let} $B = \argmin_{y \in c} \text{break}(y)$\;
        \textbf{let} $M = \argmax_{v \in B} \text{m2b2}(v)$\;
        Choose uniformly at random $x \in M$ and flip its value\;
      }
    }
    \Return{No satisfying assignment found}\;
  }
  \caption{WalkSATm2b2 Algorithm \label{walksatm2b2}}
\end{algorithm}

Intuitively, $m2b2$ rewards variables for moving clauses away from standard instability (no literals satisfied) \textit{and} for moving away from NAE instability (all literals satisfied). A similar idea, \textit{subscores}, is considered in the work of Cai \& Su \cite{cai13} where they make use of multi-level scores for $k$-SAT solvers. In this, however, $subbreak(x)$ - which considers clauses that move from being $(k-1)$-satisfied to $(k-2)$-satisfied - is \textit{minimised} since this makes $k$-SAT clauses more unstable.

In what follows, we benchmark QAOA against both \verb|WalkSATlm| and \verb|WalkSATm2b2|. For both classical solvers, we carry out a grid search across
\begin{itemize}[noitemsep, nolistsep]
    \item noise $p \in \{\frac{i}{20} : 0 \leq i \leq 20\}$
    \item weights $\omega_1 \in \{\frac{i}{10}: 0 \leq i \leq 10\}$, $\omega_2 = 1 - \omega_1$
\end{itemize} 
to identify the optimal set of hyperparameters for each $k$. In particular, in keeping with the $\u{\beta}, \u{\gamma}$ angle selection procedure for QAOA, this is done for only $t = 100$ \textbf{satisfiable} random $k$-NAE-SAT instances, $\{ \us_i : \us_i \sim CNF(\vb{n=12}, k, \hat{r}^{\text{NAE}}_k)\}_{i = 0}^{t - 1}$.

\section{Results}
For ease of exposition, we will analyse  here the results for $k = 9$ instances. The full set of results for other $k$ and $p$ are listed in Appendix
\ref{fullresults}.

\subsection{Success Probabilities}
Given parameters $\underline{\beta}^*$ and $\underline{\gamma}^*$, we evaluate the success probability of QAOA over $v = 2500$ \textbf{satisfiable} random $9$-NAE-SAT instances, $\{ \us_i : \us_i \sim CNF(n, k, \hat{r}^{\text{NAE}}_k)\}_{i = 0}^{v - 1}$ and calculate
\begin{equation}
    \hat{p}^{\text{NAE}}_{succ} = \frac{1}{v}\sum_{i = 0}^{v - 1} p^{\text{NAE}}_{succ}(\sigma_i) 
\end{equation}

\begin{figure}[h]
    \centering
    \includegraphics[width=0.4\textwidth]{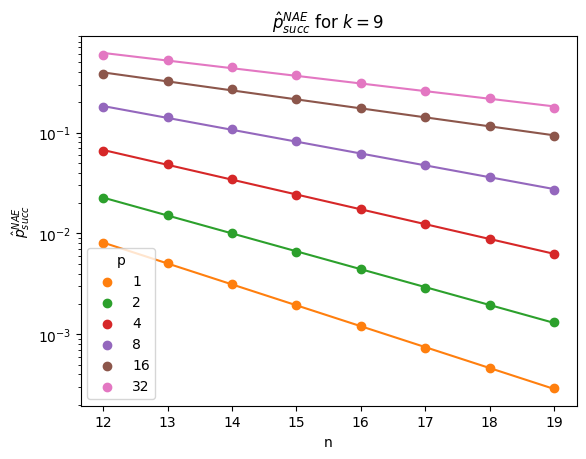}
    \vspace{-1em}
    \caption{QAOA average success probabilities across 2500 satisfiable $9$-NAE-SAT $CNF(n, 9, \hat{r}^{\text{NAE}}_{9})$ instances (error bars too small to be seen).}
    \label{fig:psucc_knaesat_k9}
\end{figure}

This is done for $p=\{1,2,4,8,16,32\}$ layers and $12 \leq n \leq 19$. The results (Figure \ref{fig:psucc_knaesat_k9}) show, as expected, an exponential decay in success probability with instance size.
The success probability increases with the number of layers $p$, while, importantly, the rate of decay decreases as $p$ increases.

\subsection{Running Times}
We define the notion of (median) running times for QAOA $k$-NAE-SAT as follows.

\bd (Instance running time) The running time of a random $k$-NAE-SAT QAOA instance, $r^{\text{NAE}}_{\us}$, is defined as the number of bitstrings that have to be sampled from the final quantum state $\ket{\Psi^{\text{NAE}}(\underline{\beta},\underline{\gamma}, \us)}$ before one is a satisfying assignment of $\us$.
\ed

\bd (Empirical median running time) Given the ensemble $R = \{ \us_i : \us_i \sim CNF(n, k, r)\}_{i = 0}^{v - 1}$, the QAOA empirical median running time $Med_{\us \in R}[r^{\text{NAE}}_{\us}]$ is defined as the median of the corresponding running times $\{ r^{\text{NAE}}_{\us_i} : \us_i \in R \}$. 
\ed

In the work of \cite{boulebnane22}, it is empirically observed that
\begin{equation}
    Med_{\us \sim CNF(n, k, r)}[r_{\us}] \sim [\mathbb{E}_{\us \sim CNF(n, k, r)} [p_{succ}(\us)]]^{-1}
\end{equation}
for $k$-SAT. We study this relationship for $k$-NAE-SAT.

\begin{figure}[h]
    \centering
    \includegraphics[width=0.39\textwidth]{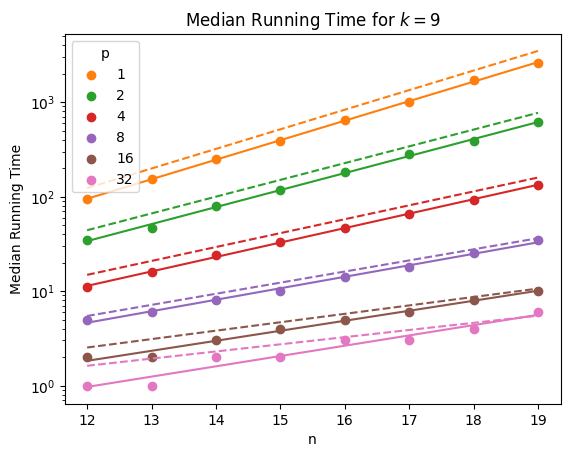}
    \vspace{-1em}
    \caption{QAOA median running times across 2500 satisfiable $9$-NAE-SAT $CNF(n, 9, \hat{r}^{\text{NAE}}_{9})$ instances. Dashed lines are the reciprocal of corresponding success probabilities.}
    \label{fig:mrt_knaesat_k9}
\end{figure}

Given parameters $\underline{\beta}^*$ and $\underline{\gamma}^*$, we evaluate the running time of QAOA over $v = 2500$ \textbf{satisfiable} random $9$-NAE-SAT instances, $R=\{ \us_i : \us_i \sim CNF(n, k, \hat{r}^{\text{NAE}})\}_{i = 0}^{v - 1}$, and calculate $Med_{\us \in R}[r^{\text{NAE}}_{\us}]$. We consider circuits with $p \in \{1, 2, 4, 8, 16, 32\}$ layers and instances where $12 \leq n \leq 19$. Our results (Figure \ref{fig:mrt_knaesat_k9}) show that the median running time scales exponentially with instance size and confirm its alignment with the reciprocal of the success probability. Importantly, the slope of scaling decreases as the number of layers $p$ increases. 

We observe (see Appendix \ref{fullresults}) that the two metrics agree more strongly for larger $k$ and hypothesise that this is related to the discrete nature of the running time in contrast to the continuity of the success probability. In particular, since a smaller $k$ corresponds to generally less challenging instances, the QAOA running times are similar across instance sizes and produce flat scalings. A similar effect occurs for larger values of $p$. In this case, we expect the QAOA circuits with more layers to be more powerful, requiring fewer samples to extract a satisfying assignment. 

The work of Boulebnane \& Montanaro \cite{boulebnane22} proves (under certain assumptions) that for fixed $p$
\begin{equation}
    \lim_{n \rightarrow \infty} \frac{1}{n} \log \mathbb{E}_{\us \sim CNF(n, k, r)}[p_{succ}(\us)]
\end{equation}
exists for $k$-SAT. Extending this, we investigate the exponents induced by the two metrics for NAE-SAT as follows.

\bd (NAE Empirical success probability scaling exponent)
Given the ensemble \\ $\{ \us_i : \us_i \sim CNF(n, k, r)\}_{i = 0}^{v - 1}$, $\hat{C}^{\text{NAE}}_{p, k}(\underline{\beta}, \underline{\gamma})$ is such that 
\begin{equation} \label{naempsucc}
     \hat{p}^{\text{NAE}}_{succ} = 2^{-n \hat{C}^{\text{NAE}}_{p, k}(\underline{\beta}, \underline{\gamma}) + o(n)}
\end{equation}
\ed

\bd (NAE Empirical median running time scaling exponent)
Given the ensemble $R = \{ \us_i : \us_i \sim CNF(n, k, r)\}_{i = 0}^{v - 1}$, $\Tilde{C}^{\text{NAE}}_{p, k}(\underline{\beta}, \underline{\gamma})$ is such that
\begin{equation} \label{naempmed}
     Med_{\us \in R}[r^{\text{NAE}}_{\us}] = 2^{n \Tilde{C}^{\text{NAE}}_{p, k}(\underline{\beta}, \underline{\gamma}) + o(n)}
\end{equation}
\ed

We consider the relative error of these exponents across $3 \leq k \leq 10$, defined as
\begin{equation}
    \left\lvert \frac{\hat{C}^{\text{NAE}}_{p, k}(\underline{\beta}, \underline{\gamma})-\Tilde{C}^{\text{NAE}}_{p, k}(\underline{\beta}, \underline{\gamma})}{\hat{C}^{\text{NAE}}_{p, k}(\underline{\beta}, \underline{\gamma})} \right\rvert
\end{equation}
where each exponent $c$ is computed by fitting to a power law of the form $2^{cn + d}$ ($d$ is a constant independent of $n$).

\begin{figure}[h]
    \centering
    \label{fig:relative_error_k_p}
    \includegraphics[width=0.5\textwidth]{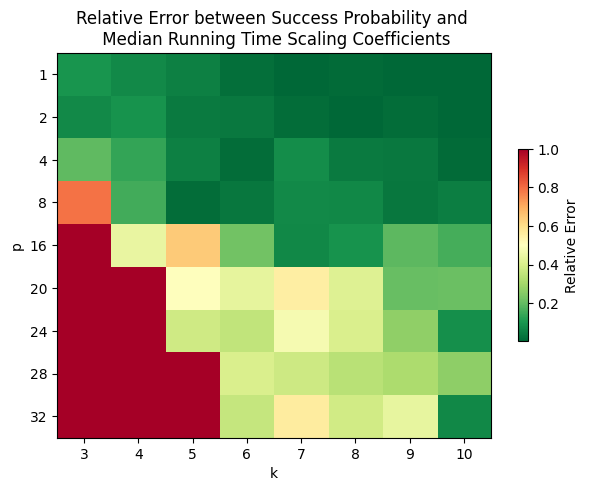}
    \vspace{-2em}
    \cprotect\caption{$\hat{C}^{\text{NAE}}_{p, k}(\underline{\beta}, \underline{\gamma})$, $\Tilde{C}^{\text{NAE}}_{p, k}(\underline{\beta}, \underline{\gamma})$ relative error across 2500 satisfiable $k$-NAE-SAT $CNF(n, k,\hat{r}^{\text{NAE}}_k)$ instances. \label{fig:relative_error}}
\end{figure}

The results (Figure \ref{fig:relative_error}) support our observations: we find that the error decreases as $k$ increases and increases as $p$ increases. We note that these effects are also in part due to the relatively small values of $n$ which we are considering. We do not expect that the median running times will continue to scale as flatly in the large $n$ limit, even for the large $p$ that we have considered. 

\subsection{Benchmarking}
We compare the median running time of QAOA to that of \verb|WalkSATlm| and \verb|WalkSATm2b2| across 2500 \textbf{satisfiable} random 9-NAE-SAT instances, $\{ \us_i : \us_i \sim CNF(n, k, \hat{r}^{\text{NAE}}_k\}_{i = 0}^{v - 1}$, for $12 \leq n \leq 19$. 

We define the run time of both classical solvers as the number of loop iterations made by the algorithm. Comparing this to QAOA's run time is justified by considering the cost of operations associated with each. For \verb|WalkSATlm|, one iteration involves
\begin{equation}
    \order{mk^2}
\end{equation}
operations to consider the $score$ of each of $k$ variables in a clause. Calculating the $score$ requires traversing each of the $k$ variables in each of the $m$ clauses. A similar argument can be applied for \verb|WalkSATm2b2|. For QAOA, one sample corresponds to measuring the state
\begin{equation}
    \ket{\Psi(\u{\beta}, \u{\gamma}, \us)}_p := \prod\limits_{i = 0}^{p-1}\underbrace{\unb(\beta_i)}_{n \text{ gates}}\underbrace{\uncsn(\gamma_i)\ket{s}}_{2mk^3 \text{ gates}}
\end{equation}
and so requires
\begin{equation}
    \order{p(mk^3 + n)}
\end{equation}
gates - where we have absorbed the $\order{n}$ cost of measurement. As $m = rn$, we deduce that for fixed $k, p$ and $r$, one iteration in both algorithms is an $\order{n}$ operation. Since we are primarily interested in the scaling of the algorithms' running times, which follow an exponential fit $\sim 2^{cn}$, the comparison is valid.

\begin{figure}[h]
    \centering
    \includegraphics[width=.5\textwidth]{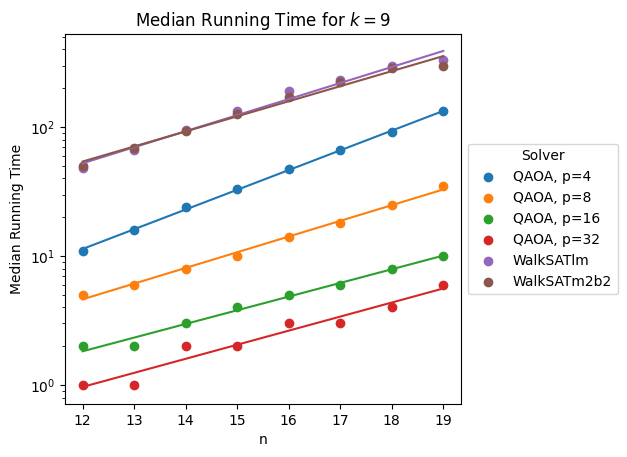}
    \vspace{-2em}
    \cprotect\caption{Median running times across 2500 satisfiable 9-NAE-SAT $CNF(n, k, \hat{r}^{\text{NAE}}_k)$ instances of QAOA, \verb|WalkSATlm| and \verb|WalkSATm2b2|.}
    \label{fig:benchmark_knaesat_k9}
\end{figure}

We observe (Figure \ref{fig:benchmark_knaesat_k9}) that the median running time of QAOA outperforms \verb|WalkSATm2b2| which in turn outperforms \verb|WalkSATlm|. Further, we assess the scaling of these runtimes rather than their absolute values (Figure \ref{fig:knaesat-excessive-scaling}) and find that there exists a threshold $p$ above which QAOA's scaling improves on \verb|WalkSATm2b2|. We include the results for $k=5$ for comparison, fit the coefficients to a power law $\sim ap^b$ and calculate that the coefficients scale as
\begin{equation}
    \Tilde{C}^{\text{NAE}}_{p, 5}(\underline{\beta}, \underline{\gamma}) \sim 0.64 p^{-0.52}
\end{equation}
\begin{equation}
    \Tilde{C}^{\text{NAE}}_{p, 9}(\underline{\beta}, \underline{\gamma}) \sim 0.69 p^{-0.22}
\end{equation}

The power law fit for $k=5$ begins to disagree with the computed coefficients in the large $p$ limit. As in the previous section, this is likely due to the fact that the instances are easy for QAOA to solve for the limited values of the problem size $n$ we considered. As such, effectively no scaling is induced since the running times are generally $\approx 1$. As a result, we emphasise that the coefficients computed in this regime should be investigated further through experiments on instances with larger values of $n$. It is clear that this is no longer the case for $k=9$ where we find improved agreement due to harder problem instances. 

We include (Table \ref{crossover}) the values of $p$ for which QAOA's scaling begins to outperform that of \verb|WalkSATm2b2| (see Appendix \ref{fullresults} for details of these results). We also identify that QAOA outperforms Grover's search algorithm \cite{grover1996fast}, which induces an exponent of $1/2$, for respective values of $p$.   

\begin{table}[h!]
\centering
\begin{tabular}{|c|ccccccc|}
\hline
$k$ & 3 & 4 & 5 & 6 & 7 & 8 & 9 \\
\hline
$p$ & 3 & 3 & 4 & 6 & 6 & 10 & 14 \\
\hline
\end{tabular}
\cprotect\caption{Value of $p$ for which QAOA's scaling $ \Tilde{C}^{\text{NAE}}_{p, k} (\underline{\beta}, \underline{\gamma})$ outperforms \verb|WalkSATm2b2| on $k$-NAE-SAT. \label{crossover}}
\end{table}

\subsection{Excessive Scaling}
We examine the notion of \textit{excessive scaling}: the scaling induced by QAOA in cases where the initial state $\ket{+}^{\otimes n}$ is mostly unchanged by the circuit.

\bd (NAE Predicted scaling exponent)
\begin{equation}
C^{\text{NAE}}_{p, k} (\underline{\beta}, \underline{\gamma}) = -\lim_{n \rightarrow \infty} \frac{1}{n} \log \mathbb{E}_{\us \sim CNF(n, k, r)}[p^{\text{NAE}}_{succ}(\us)] 
\end{equation}
\ed

We show (see Appendix \ref{exscal}) that
\begin{equation}
C^{\text{NAE}}_{p, k} (\underline{\beta}, \underline{0}) = C^{\text{NAE}}_{p, k} (\underline{0}, \underline{\gamma}) = 2^{1-k}r
\end{equation}

\begin{figure}[h]
    \centering
    \includegraphics[width=0.5\textwidth]{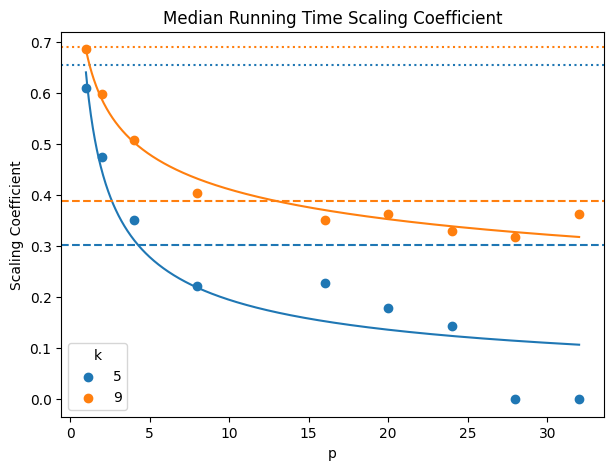}
    \vspace{-2em}
    \cprotect\caption{Induced scaling of QAOA $k$-NAE-SAT median running times $\Tilde{C}^{\text{NAE}}_{p, k}(\underline{\beta}, \underline{\gamma})$. Dashed line is observed \verb|WalkSATm2b2| scaling, dotted line is random assignment scaling and solid lines are the power law fits that induce $\Tilde{C}^{\text{NAE}}_{p, k}(\underline{\beta}, \underline{\gamma})$. Points for large $p$ at $k=5$ suffer from strong finite size effects.}
    \label{fig:knaesat-excessive-scaling}
\end{figure}

As the size of the problem ensemble increases, we expect $\hat{C}^{\text{NAE}}_{p, k}(\underline{\beta}, \underline{\gamma}) \approx C^{\text{NAE}}_{p, k} (\underline{\beta}, \underline{\gamma})$ (\ref{naempsucc}). Further, we have studied the relationship between $\hat{C}^{\text{NAE}}_{p, k}(\underline{\beta}, \underline{\gamma})$ and $\Tilde{C}^{\text{NAE}}_{p, k}(\underline{\beta}, \underline{\gamma})$ (Figure \ref{fig:relative_error}). As such, we compare $2^{1-k}r$ to the empirical observations $\Tilde{C}^{\text{NAE}}_{p, k}(\underline{\beta}, \underline{\gamma})$ in our experiments. As anticipated, QAOA outperforms the random scaling (Figure \ref{fig:knaesat-excessive-scaling}). We observe that in the small $p$ regime, the scaling is effectively that of random assignment. This is due to the fact that few-layer circuits are inexpressive leading to outputs that are effectively unchanged from the initial state.

\section{Conclusions and Future Work}
The results above provide three main contributions, in addition to the QAOA encodings derived for $k$-NAE-SAT.

First, we have shown that the success probability agrees strongly with the median running time. This suggests that we can make use of an analytic heuristic to predict the empirical performance of QAOA, over an ensemble of random instances. This is particularly useful in the current NISQ-era, where we only have access to quantum processors with low gate fidelities and high qubit decoherence rates \cite{sun23}. Meanwhile, classical simulations are limited to less than $50$ qubits \cite{wu2019} so don't provide a practical means of predicting the large $n$ QAOA performance. We explored the strength of agreement between the success probability and median running time and identified regimes where they diverge. However, these were primarily related to problem instances being too easy or the QAOA circuits too powerful - both situations of lesser interest for the purposes of solving difficult problems.

We have introduced a novel stochastic local search algorithm for $k$-NAE-SAT, \verb|WalkSATm2b2|, and shown that it outperforms its predecessor \verb|WalkSATlm|. This identifies that the symmetries of the problem can be exploited to improve running time. Benchmarking QAOA against both solvers, we find a quantum advantage across all $k$-NAE-SAT instances of $3 \leq k \leq 9$. Importantly, all solver hyperparameters were fine-tuned over the same set of problem instances, suggesting that QAOA is more capable of generalising.

Finally, we considered the excessive scaling of QAOA for $k$-NAE-SAT and showed that for small $p$, QAOA circuits act closely to a random assignment algorithm. This confirms the inexpressivity of shallow depth circuits. Nonetheless, the scaling rapidly decreases for a small increase in the number of layers, highlighting the effectiveness of the QAOA ansatz and suggesting that large circuits are not necessary for advantages over classical algorithms.

Interesting future directions are deriving analytical formulas for the success probability in the large $n$ limit similar to the case of $k$-SAT  \cite{boulebnane22}, and benchmarking QAOA against classical solvers for hard instances of other random constrained satisfaction problems \cite{coja09}.

\section*{Acknowledgements}
The authors acknowledge helpful discussions with Sami Boulebnane.

\section*{Code}
Code is available from \url{https://github.com/Quantum-AI-Lab-ICL/QAOA-SAT}.

\bibliography{main}
\bibliographystyle{unsrt}

\onecolumn\newpage
\appendix
\section{Gate Based QAOA Implementation} \label{qaoagatebased}
\subsection{Representing Functions as Hamiltonians}
We consider the notion of \textit{representing Hamiltonians} as follows.
\bd \label{frep}
(Hamiltonian representation of a function) A Hamiltonian $\ham_f$ represents a function $f$ if:
\begin{equation}
    \ham_f\ketx = f(\ux)\ketx
\end{equation}

for $\ux \in \bn^n$ with corresponding computational basis state $\ketx$ \cite{hadfield21}.
\ed

\subsubsection{Boolean Functions}
\bd
(Boolean function) The class of Boolean function on $n$ bits takes the form 
\begin{equation}
\mathcal{B}_n := \{ f: \bn^n \rightarrow \bn \}
\end{equation}
\ed

We begin by representing $f(\ux) = 1$ by $\ham_f = \mathbb{I}$ and $g(\ux) = 0$ by $\ham_g = 0$, where $\mathbb{I}$ is the identity operator and $0$ the null operator.

It is clear that
\begin{equation}
    \begin{split}
        \ham_f \ketx &= 1 \times \ketx = \ketx\\
        \ham_g \ketx &= 0 \times \ketx = 0
    \end{split}
\end{equation}

Next, we represent $h_j(\ux) = x_j$ where $\ux = x_0x_1\dots x_{n-1}$, making use of the 
Pauli-Z operator $Z = \ket{0}\bra{0} - \ket{1}\bra{1}$. We consider the multiple qubit version 
\begin{equation}
    Z_j := I^{\otimes(j - 1)} \otimes Z \otimes I^{\otimes(n - j)}
\end{equation}
whose application gives
\begin{equation}
    Z_j \ketx = (-1)^{x_j}\ketx = (1 - 2x_j)\ketx
\end{equation}
Letting $\ham_{h_j} := \frac{1}{2}(\mathbb{I} - Z_j)\ketx$
\begin{equation} \label{extractor}
    \ham_{h_j} \ketx = \frac{1}{2}(1 - (1 - 2x_j))\ketx = x_j\ketx
\end{equation}

It is clear $\ham_{h_j}$ represents $h_j$. Combining these together, the following composition rules can be derived. 

\bt \label{comprules}
Let $f, g \in \mathcal{B}_n$ represented by $\ham_f, \ham_g$, then \cite{hadfield21}:
\begin{itemize}
    \item $\ham_{\neg f} = \mathbb{I} - \ham_f$
    \item $\ham_{f \land g} = \ham_f\ham_g$
    \item $\ham_{f \lor g} = \ham_f + \ham_g - \ham_f\ham_g$
    \item $\ham_{f \oplus g} = \ham_f + \ham_g -2\ham_f\ham_g$
    \item $\ham_{f \Rightarrow g} = \mathbb{I} - \ham_f + \ham_f\ham_g$
    \item $\ham_{af + bg} = a\ham_f + b\ham_g$
\end{itemize}
\et

\subsubsection{Real/Pseudo-Boolean Functions}
In many cases, functions of interest take the form of a (weighted) sum of Boolean functions

\bd
(Real function) The class of Real functions on $n$ bits take the form 
\begin{equation}
\mathcal{R}_n := \{ f: \bn^n \rightarrow \mathbb{R} \}
\end{equation}
\ed

\bt \label{repreal}
Every $g \in \mathcal{R}_n$ can be written (non-uniquely) as the weighted sum of Boolean functions \cite{hadfield21}:
\begin{equation}
g(\ux) = \sum\limits_{j} w_jf_j(\ux)
\end{equation}
where $f_j \in \mathcal{B}_{m \leq n}$ (i.e. acts on a subset of the bits) and $w_j \in \mathbb{R}$.
\et

Taking $\mathcal{B}_n$ as the elements of a real vector space, they form a basis of $\mathcal{R}_n$ for each $n$. As such, the terms $f_j, w_j$ above are effectively basis vectors and their corresponding coefficients in the expansion of the given Real function $g$. A Real function in this representation is termed \textit{Pseudo-Boolean}.

\bt \label{pseudof}
Let $g \in \mathcal{R}_n$ a Pseudo-Boolean function, its representing Hamiltonian takes the form \cite{hadfield21}:
\begin{equation}
\ham_g = \sum\limits_{j} w_j\ham_{f_j}
\end{equation}
where $\ham_{f_j}$ is the representing Hamiltonian of $f_j \in \mathcal{B}_n$. 
\et

\subsection{Problem Hamiltonian}
Our objective function to be minimised is
\begin{equation}
\mathcal{C}^{\text{NAE}}_{\us}(\ux) = \sum\limits_{i = 0}^{m-1} \mathbbm{1}\{\ux \nvdash_{\text{NAE}} \sigma_i\}
\end{equation}

We note however that
\begin{equation}
    \ux \vdash_{\text{NAE}} \sigma_i \equiv \left[ \ux \vdash \sigma_i \wedge \ux \nvdash \bigwedge_{j = 0}^{k-1} l_{\sigma_{ij}} \right]
\end{equation}
So, by De Morgan's law 
\begin{equation}\label{demorgannae}
    \ux \nvdash_{\text{NAE}} \sigma_i \equiv \left[\ux \nvdash \sigma_i \vee \ux \vdash \bigwedge_{j = 0}^{k-1} l_{\sigma_{ij}}\right]
\end{equation}
In other words, the assignment is penalised if it does not satisfy the clause or if it sets all its literals to true. As such, we can write

\begin{equation} \label{knaesatobj}
    \mathcal{C}^{\text{NAE}}_{\us}(\ux) = \sum\limits_{i = 0}^{m-1} \mathbbm{1}\{\ux \nvdash_{\text{NAE}} \sigma_i\} = \sum\limits_{i = 0}^{m-1} \mathbbm{1}\{\ux \nvdash \sigma_i \vee \ux \vdash \bigwedge_{j = 0}^{k-1} l_{\sigma_{ij}}\} 
\end{equation}

We identify $\mathcal{C}^{\text{NAE}}_{\us}$ as a Pseudo-Boolean function. Its representing Hamiltonian takes the form
\begin{equation} \label{knaerep}
\ham_{\mathcal{C}^{\text{NAE}}_{\us}} = \sum\limits_{i = 0}^{m-1} \ham_{\mathbbm{1}\{\ux \nvdash \sigma_i \vee \ux \vdash \bigwedge_{j = 0}^{k-1} l_{\sigma_{ij}}\}} \equiv \sum\limits_{i = 0}^{m-1} \ham_{\neg \sigma_i \vee \left(\bigwedge_{j = 0}^{k-1} l_{\sigma_{ij}}\right)}
\end{equation}

where we have treated the clauses and literals as Boolean functions.
\begin{equation}
    \sigma_i : \bn^n \rightarrow \bn
\end{equation}
\begin{equation}
    l_{ij} : \bn \rightarrow \bn
\end{equation}
such that
\begin{equation}
    \sigma_i(\ux) = 
    \begin{cases}
         1 & \ux \vdash \sigma_i \\
         0 & \ux \nvdash \sigma_i
    \end{cases}
\end{equation}
\begin{equation}
    l_{ij}(x) = 
    \begin{cases}
         1 & x \vdash l_{ij} \\
         0 & x \nvdash l_{ij}
    \end{cases}
\end{equation}

Applying composition rules
\begin{equation}
    \ham_{\neg \sigma_i \vee \left(\bigwedge_{j = 0}^{k-1} l_{\sigma_{ij}}\right)} = \ham_{\neg \sigma_i} + \ham_{\bigwedge_{j = 0}^{k-1} l_{\sigma_{ij}}} - \ham_{\neg \sigma_i}\ham_{\bigwedge_{j = 0}^{k-1} l_{\sigma_{ij}}}
\end{equation}

It is clear that the third Hamiltonian, $\ham_{\neg \sigma_i}\ham_{\bigwedge_{j = 0}^{k-1} l_{\sigma_{ij}}}$  vanishes, namely $\ux$ cannot both satisfy $\bigwedge_{j = 0}^{k-1} l_{\sigma_{ij}}$ while not satisfying $\sigma_i = \bigvee_{j = 0}^{k-1} l_{\sigma_{ij}}$.

By De Morgan's law, the first Hamiltonian expands to give
\begin{equation}
    \ham_{\neg \sigma_i} = \ham_{\neg \left(\bigvee_{j = 0}^{k-1} l_{\sigma_{ij}}\right)} \equiv \ham_{\bigwedge_{j = 0}^{k-1} \neg l_{\sigma_{ij}}}
\end{equation}

Applying composition laws
\begin{equation}
\hat{\mathcal{H}}_{\bigwedge_{j = 0}^{k-1} \neg l_{\sigma_{ij}}} = \prod_{j = 0}^{k-1}\hat{\mathcal{H}}_{\neg l_{\sigma_{ij}}} = \prod_{j = 0}^{k-1}(\mathbb{I} - \hat{\mathcal{H}}_{l_{\sigma_{ij}}})
\end{equation}

$\ham_{f_i} = \frac{1}{2}(\mathbb{I} - Z_i)$ represents $f_i(\ux) = x_i$ and $\ham_{f_i} = \frac{1}{2}(\mathbb{I} + Z_i)$ represents $f_i(\ux) = \neg x_i$, we combine these to derive that:

\begin{equation} \label{repliteral}
    \ham_{l_{\sigma_{ij}}} = \frac{1}{2}\mathbb{I} + s_{\sigma_{ij}}\frac{1}{2}Z_{\sigma_{ij}}
\end{equation}

represents $l_{\sigma_{ij}}$, where $s_{\sigma_{ij}} = -1$ for positive literals $l_{\sigma_{ij}} = x_{\sigma_{ij}}$ and $s_{\sigma_{ij}} = 1$ for negative literals $l_{\sigma_{ij}} = \neg x_{\sigma_{ij}}$. Putting this together, we find:

\begin{equation}
\hat{\mathcal{H}}_{\neg \sigma_i} = \prod_{j = 0}^{k-1}\left(\mathbb{I} - \hat{\mathcal{H}}_{l_{\sigma_{ij}}}\right) = \prod_{j = 0}^{k-1}\left(\mathbb{I} - \frac{1}{2}\mathbb{I} - s_{\sigma_{ij}}\frac{1}{2}Z_{\sigma_{ij}}\right) = \frac{1}{2^k}\prod_{j = 0}^{k-1}\left( \mathbb{I} - s_{\sigma_{ij}}Z_{\sigma_{ij}} \right)
\end{equation}

Expanding, and ignoring constants as this is a function being minimised, we arrive at:
\begin{equation} \label{hnsigma}
    \hat{\mathcal{H}}_{\neg \sigma_i} = \frac{1}{2^k}\left(-\sum_{a = 0}^{k-1} s_{\sigma_{ia}}Z_{\sigma_{ia}} + \sum_{b > a}^{k-1} s_{\sigma_{ia}}s_{\sigma_{ib}}Z_{\sigma_{ia}}Z_{\sigma_{ib}} - \sum_{c > b > a}^{k-1} s_{\sigma_{ia}}s_{\sigma_{ib}}s_{\sigma_{ic}}Z_{\sigma_{ia}}Z_{\sigma_{ib}}Z_{\sigma_{ic}} \dots \right)
\end{equation}

Finally, applying composition rules and the representation of a literal, the middle Hamiltonian takes the form
\begin{equation}
    \ham_{\bigwedge_{j = 0}^{k-1} l_{\sigma_{ij}}} = \prod_{j = 0}^{k-1}\hat{\mathcal{H}}_{l_{\sigma_{ij}}} = \prod_{j = 0}^{k-1}\left(\frac{1}{2}\mathbb{I} + s_{\sigma_{ij}}\frac{1}{2}Z_{\sigma_{ij}}\right) = \frac{1}{2^k}\prod_{j = 0}^{k-1}\left( \mathbb{I} + s_{\sigma_{ij}}Z_{\sigma_{ij}} \right)
\end{equation}

Expanding, and ignoring constants as this is a function being minimised:
\begin{equation}
    \ham_{\bigwedge_{j = 0}^{k-1} l_{\sigma_{ij}}} = \frac{1}{2^k}\left(\sum_{a = 0}^{k-1} s_{\sigma_{ia}}Z_{\sigma_{ia}} + \sum_{b > a}^{k-1} s_{\sigma_{ia}}s_{\sigma_{ib}}Z_{\sigma_{ia}}Z_{\sigma_{ib}} + \sum_{c > b > a}^{k-1} s_{\sigma_{ia}}s_{\sigma_{ib}}s_{\sigma_{ic}}Z_{\sigma_{ia}}Z_{\sigma_{ib}}Z_{\sigma_{ic}} \dots \right)
\end{equation}

As such, the unitary operator takes the form
\begin{equation}
\begin{split}
	\uncsn(\gamma) = \exp\left[-i \gamma \hamcsn \right] &= \exp \left[-i \gamma \sum_{j = 0}^{m-1} \ham_{\neg \sigma_j} + \ham_{\bigwedge_{q = 0}^{k-1} l_{\sigma_{jq}}}\right]\\
    &= \prod_{j = 0}^{m-1} \exp \left[-i \gamma \left(\hat{\mathcal{H}}_{\neg \sigma_j}+ \ham_{\bigwedge_{q = 0}^{k-1} l_{\sigma_{jq}}}\right) \right] \\
    &= \prod_{j = 0}^{m-1} \exp \left[-i \gamma \hat{\mathcal{H}}_{\neg \sigma_j}\right] \exp\left[-i \gamma \ham_{\bigwedge_{q = 0}^{k-1} l_{\sigma_{jq}}}\right] \\
    &:= \prod_{j = 0}^{m-1} \hat{\mathcal{U}}_{\neg \sigma_j}(\gamma)\hat{\mathcal{U}}_{\bigwedge_{q = 0}^{k-1} l_{\sigma_{jq}}}(\gamma)
\end{split}
\end{equation}

Where we have used the fact that both Hamiltonians only consist of $\mathbbm{1}$ and Pauli $Z$ operators, meaning they commute. 
\begin{equation} \label{unsigma}
\begin{split}
	&\hat{\mathcal{U}}_{\neg \sigma_j}(\gamma)\\
    &= \exp \left[-i \gamma \frac{1}{2^k}\left(-\sum_{a = 0}^{k-1} \theta_{a}Z_{\sigma_{ja}} + \sum_{b > a}^{k-1} \theta_{ab}Z_{\sigma_{ja}}Z_{\sigma_{jb}} - \sum_{c > b > a}^{k-1} \theta_{abc}Z_{\sigma_{ja}}Z_{\sigma_{jb}}Z_{\sigma_{jc}} \dots \right) \right] \\
	&= \exp \left[i \gamma \frac{1}{2^k}\sum_{a = 0}^{k-1} \theta_{a}Z_{\sigma_{ja}} \right] \exp \left[-i \gamma \frac{1}{2^k} \sum_{b > a}^{k-1} \theta_{ab}Z_{\sigma_{ja}}Z_{\sigma_{jb}} \right] \exp \left[i \gamma \frac{1}{2^k} \sum_{c > b > a}^{k-1} \theta_{abc}Z_{\sigma_{ja}}Z_{\sigma_{jb}}Z_{\sigma_{jc}} \right] \dots \\
	&= \prod_{a = 0}^{k-1} \exp \left[i \gamma \frac{1}{2^k} \theta_{a}Z_{\sigma_{ja}} \right] \prod_{b > a}^{k-1}\exp \left[-i \gamma \frac{1}{2^k}  \theta_{ab}Z_{\sigma_{ja}}Z_{\sigma_{jb}} \right] \prod_{c > b > a}^{k-1}\exp \left[i \gamma \frac{1}{2^k}  \theta_{abc}Z_{\sigma_{ja}}Z_{\sigma_{jb}}Z_{\sigma_{jc}} \right] \dots \\
	&= \prod_{a = 0}^{k-1} R_{Z_{\sigma_{ja}}}\left[-\frac{\gamma}{2^{k-1}} \theta_{a} \right] \prod_{b > a}^{k-1}R_{Z_{\sigma_{ja}}Z_{\sigma_{jb}}} \left[\frac{\gamma}{2^{k-1}}  \theta_{ab} \right] \prod_{c > b > a}^{k-1}R_{Z_{\sigma_{ja}}Z_{\sigma_{jb}}Z_{\sigma_{jc}}} \left[-\frac{\gamma}{2^{k-1}}  \theta_{abc} \right] \dots \\
\end{split}
\end{equation}

while
\begin{equation}
\begin{split}
	&\hat{\mathcal{U}}_{\bigwedge_{q = 0}^{k-1} l_{\sigma_{jq}}}(\gamma)\\
    &= \exp \left[-i \gamma \frac{1}{2^k}\left(\sum_{a = 0}^{k-1} \theta_{a}Z_{\sigma_{ja}} + \sum_{b > a}^{k-1} \theta_{ab}Z_{\sigma_{ja}}Z_{\sigma_{jb}} + \sum_{c > b > a}^{k-1} \theta_{abc}Z_{\sigma_{ja}}Z_{\sigma_{jb}}Z_{\sigma_{jc}} \dots \right) \right] \\
	&= \exp \left[-i \gamma \frac{1}{2^k}\sum_{a = 0}^{k-1} \theta_{a}Z_{\sigma_{ja}} \right] \exp \left[-i \gamma \frac{1}{2^k} \sum_{b > a}^{k-1} \theta_{ab}Z_{\sigma_{ja}}Z_{\sigma_{jb}} \right] \exp \left[-i \gamma \frac{1}{2^k} \sum_{c > b > a}^{k-1} \theta_{abc}Z_{\sigma_{ja}}Z_{\sigma_{jb}}Z_{\sigma_{jc}} \right] \dots \\
	&= \prod_{a = 0}^{k-1} \exp \left[-i \gamma \frac{1}{2^k} \theta_{a}Z_{\sigma_{ja}} \right] \prod_{b > a}^{k-1}\exp \left[-i \gamma \frac{1}{2^k}  \theta_{ab}Z_{\sigma_{ja}}Z_{\sigma_{jb}} \right] \prod_{c > b > a}^{k-1}\exp \left[-i \gamma \frac{1}{2^k}  \theta_{abc}Z_{\sigma_{ja}}Z_{\sigma_{jb}}Z_{\sigma_{jc}} \right] \dots \\
	&= \prod_{a = 0}^{k-1} R_{Z_{\sigma_{ja}}}\left[\frac{\gamma}{2^{k-1}} \theta_{a} \right] \prod_{b > a}^{k-1}R_{Z_{\sigma_{ja}}Z_{\sigma_{jb}}} \left[\frac{\gamma}{2^{k-1}}  \theta_{ab} \right] \prod_{c > b > a}^{k-1}R_{Z_{\sigma_{ja}}Z_{\sigma_{jb}}Z_{\sigma_{jc}}} \left[\frac{\gamma}{2^{k-1}}  \theta_{abc} \right] \dots \\
\end{split}
\end{equation}
where 
\begin{itemize}
    \item $\theta_{ab\dots q} = s_{\sigma_{ja}}s_{\sigma_{jb}}\dots s_{\sigma_{jq}}$
    \item $R_{Z_1, Z_2, \dots, Z_l}(2\gamma)$ corresponds to the operation $\exp \left[-i \gamma \prod_{j = 1}^{l} Z_j \right]$ and is implemented using the decomposition
\begin{equation} \label{decomp}
    \exp \left[-i \gamma \prod_{j = 1}^{l} Z_j \right] = \exp \left[-i \gamma Z^{\otimes l} \right] = \left(\prod_{i = 1}^{l - 1}\text{CX}_{l - i, l - i + 1}\right)R_{Z_{l}}(2\gamma)\left(\prod_{i = 1}^{l - 1}\text{CX}_{i, i+1}\right)
\end{equation}
where CX$_{a,b}$ is the controlled-not gate with control qubit $a$ and target qubit $b$.
\end{itemize}

\begin{figure}[h]
    \centering
    \includegraphics[width=0.45\textwidth]{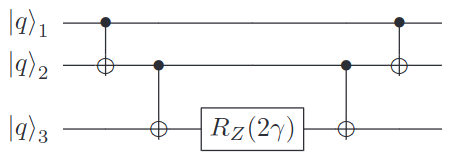}
    \cprotect\caption{Quantum circuit performing the operation $\un = \exp \left[-i \gamma Z_1Z_2Z_3 \right]$ \cite{hadfield21}.}
    \label{fig:rzzz-decomp}
\end{figure}

\subsection{Mixer Hamiltonian}
The mixer Hamiltonian is as defined as
\begin{equation}
    \hamb = \sum\limits_{j = 0}^{n-1} X_j
\end{equation}
where $X_j$ is the multi-qubit Pauli-X operator acting on the $j^{th}$ qubit.

The corresponding unitary operator is therefore
\begin{equation}
\unb(\beta) = e^{i\beta \hamb} = e^{i\beta \sum\limits_{j = 0}^{n-1} X_j } = \prod\limits_{j = 0}^{n-1} e^{i\beta X_j} = \prod\limits_{j = 0}^{n-1} R_{X_j}(-2\beta)
\end{equation}
where $R_{X_j}(\theta)$ is the operator corresponding to a rotation of the $j^{th}$ qubit by $\theta$ about the x-axis on the Bloch sphere. We note the rotation sign due to the problem being recast as a minimisation.

\subsection{Efficient Classical Simulation} \label{qaoacost}
For a system size of $N = 2^n$, states are classically represented as vectors, $z \in \mathbb{C}^N$, and operators as matrices, $\hat{\mathcal{U}} \in \mathbb{C}^{N \times N}$. As such, the application of a gate can be realised as matrix-vector multiplication, requiring $\order{N^2}$ operations. Therefore, we deduce that naively
\begin{equation}
	\un_{\neg \sigma_i} = \overbrace{\prod_{a = 0}^{k-1} \underbrace{R_{Z_{\sigma_{ia}}}(\theta_a)}_{N^2}}^k \overbrace{\prod_{b > a}^{k-1}\underbrace{R_{Z_{\sigma_{ia}}Z_{\sigma_{ib}}} (\theta_{ab})}_{3N^2}}^{k - 1} \overbrace{\prod_{c > b > a}^{k-1}\underbrace{R_{Z_{\sigma_{ia}}Z_{\sigma_{ib}}Z_{\sigma_{ic}}}(\theta_{abc})}_{5N^2}}^{k - 2} \dots \\
\end{equation}
requires
\begin{equation}
    N^2 \sum_{l = 0}^k (k - l + 1)(2l - 1) = N^2\frac{1}{6}k(k+1)(2k+1)
\end{equation}
operations, where we have used that the decomposition of $R_{Z_1, Z_2, \dots, Z_l}(\theta)$ consists of $2(l - 1) + 1$ gates. A similar argument can be applied for $\hat{\mathcal{U}}_{\bigwedge_{q = 0}^{k-1} l_{\sigma_{jq}}}(\gamma)$. Thus, the application of $\uncsn$ requires:
\begin{equation} \label{uncoperations}
    2mN^2\frac{1}{6}k(k+1)(2k+1) = \order{mN^2k^3} = \order{rk^3N^2\log N}
\end{equation}
operations since $m = rn$ and $n = \log N$. Near the satisfiability threshold the clause density takes the value $r = r_k \sim 2^{k-1} \ln{2}$ and so we find the application of $\uncsn$ requires
\begin{equation}
    \order{2^k k^3 N^2\log N}
\end{equation}
operations. In the large $n$ limit, $N^2$ dominates, giving a cost of $\order{N^2}$. However, for the purposes of simulation and, in particular, for cases where $n \sim k$, the pre-factors are computationally relevant. In addition, optimising parameters over the circuit means inefficiencies are aggregated during the automatic differentiation procedure. 

Considering the diagonalisation
\begin{equation}
    \uncsn(\gamma) = e^{-i\gamma\hamcsn} = \sumb e^{-i\gamma \mathcal{C}^{\text{NAE}}_{\us}(\ux)} \ket{\ux}\bra{\ux}
\end{equation}

Or, in matrix form
\begin{equation}
    \text{diag}\left[\csn{0}, \csn{1}, \dots, \csn{N-1}\right]
\end{equation}
where we have written $\ux$ in its decimal representation
\begin{equation} \label{decrep}
    \csn{\ux} \equiv \mathcal{C}^{\text{NAE}}_{\us}\left(\sum_{l=0}^{n-1} x_l 2^{-l}\right)
\end{equation}

Applying a diagonal matrix only requires $\order{N}$ operations. The calculation of $\csn{\ux}$ requires traversing $m = r\log N \sim 2^k \log{N}$ clauses and $k$ literals per clause. As such, the application of this diagonalisation only requires
\begin{equation}
    \order{2^k k N\log{N}}
\end{equation}
operations.

Similarly, the naive application of 
\begin{equation}
    \unb(\beta) = \overbrace{\prod\limits_{j = 0}^{n-1} \underbrace{R_{X_j}(-2\beta)}_{N^2}}^{n}
\end{equation}
requires $\order{N^2\log N}$ operations, where $n = \log N$. However, writing
\begin{equation}
    \unb(\beta) = \prod\limits_{j = 0}^{n-1} e^{i\beta X_j} = \prod\limits_{j = 0}^{n -1}\left[\cos(\beta)\mathbb{I} + i\sin(\beta)X_j\right]
\end{equation}
we realise that $\unb$ can be implemented in $\order{N \log N}$ operations. First,
\begin{equation}
\begin{split}
\left[\cos(\beta)\mathbb{I} + i\sin(\beta)X_j\right] \ketx &= \cos(\beta)\underbrace{\mathbb{I}\ket{x_0 \dots x_{n-1}}}_{0 \text{ operations}} + i\sin(\beta)\underbrace{X_j\ket{x_0 \dots x_{n-1}}}_{\order{1} \text{ operations}}\\
&= \underbrace{\cos(\beta)\ket{x_0 \dots x_{n-1}}}_{N \text{ operations}} + \underbrace{i\sin(\beta)\ket{x_0 \dots \bar{x}_{j-1} \dots x_{n-1}}}_{N \text{ operations}}
\end{split}
\end{equation}
where $\bar{x}$ denotes the flipped value of $x$ ($\bar{0} = 1, \bar{1} = 0$) and we identify that we can apply
\begin{itemize}
    \item $X_j$ in $\order{1}$ operations by realising that
    \begin{equation}
        X_j \ket{x_0 \dots x_{n-1}} = \ket{x_0 \dots \bar{x}_{j-1} \dots x_{n-1}} \equiv X_j \ketx = \ket{2^{n-j} - \ux}
    \end{equation}
    where we have again written $\ux$ in its decimal representation (\ref{decrep}). As such, applying the operator corresponds to swapping 2 elements in the corresponding vector $z \in \mathbb{C}^N$.
    \item $\cos(\beta), \sin(\beta)$ in $N$ operations by a component wise multiplication of the vector $z \in \mathbb{C}^N$.
\end{itemize}
Therefore
\begin{equation}
    \unb(\beta) = \overbrace{\prod\limits_{j = 0}^{n -1}\underbrace{\left[\cos(\beta)\mathbb{I} + i\sin(\beta)X_j\right]}_{\order{N}}}^n
\end{equation}
requires $\order{N \log N}$ operations, where $n = \log N$.\\

We emphasise that this is only a valid procedure within classical simulations. It is not feasible to directly encode such a diagonal unitary on a gate based quantum computer. Nonetheless, this does not pose an issue for success probability or run time analysis. In particular, the output of both encodings (considered as some $z \in \mathbb{C}^N$) is identical. Therefore, since the procedure is based on sampling from the output state, all calculated metrics, including the induced scaling exponents, will be unaffected.

\subsection{Recasting NAE-SAT as SAT} \label{naesatassat}
\bl (Satisfying a $k$-NAE-SAT clause) For $\sigma_i = \bigvee_{j = 0}^{k-1} l_{\sigma_{ij}}$, $\ux \vdash_{NAE} \sigma_i$ iff $\ux \vdash \sigma_i$ \textbf{and} $\bar{\ux} \vdash \sigma_i$, where $\bar{\ux}$ is the flipped assignment: $\bar{0} = 1, \bar{1} = 0$.
\begin{proof}
\begin{enumerate}
    \item["$\Rightarrow$"] Suppose $\ux \vdash_{\text{NAE}} \sigma_i $
    \begin{equation}
        \begin{split}
        &\Rightarrow \exists l_t, l_f \in \sigma_i: \ux \vdash l_t, \ux \nvdash l_f \\
        &\Rightarrow \ux \vdash \sigma_i
        \end{split}
    \end{equation}
    But $\ux \nvdash l_f \Rightarrow \bar{\ux} \vdash l_f \Rightarrow \bar{\ux} \vdash \sigma_i$.
    \item["$\Leftarrow$"] Suppose $\ux \vdash \sigma_i \text{ and } \bar{\ux} \vdash \sigma_i$
        \begin{equation}
        \begin{split}
        &\Rightarrow \exists l_t \neq l_f \in \sigma_i: \ux \vdash l_t, \bar{\ux} \vdash l_f  \\
        &\Rightarrow \exists l_t \neq l_f \in \sigma_i: \ux \vdash l_t, \ux \nvdash l_f \\
        &\Rightarrow \ux \vdash_{\text{NAE}} \sigma_i
        \end{split}
    \end{equation}
\end{enumerate}
\end{proof}
\el

Consequently
\begin{equation}
    \begin{split}
    \ux \nvdash_{\text{NAE}} \sigma_i &\iff \neg(\ux \vdash \sigma_i \wedge \bar{\ux} \vdash \sigma_i)\\
    &\iff \ux \nvdash \sigma_i \vee \bar{\ux} \nvdash \sigma_i
    \end{split}
\end{equation}
by De Morgan's law.

Similarly, the naive application of 
\begin{equation}
    \unb(\beta) = \overbrace{\prod\limits_{j = 0}^{n-1} \underbrace{R_{X_j}(-2\beta)}_{N^2}}^{n}
\end{equation}
requires $\order{N^2\log N}$ operations, where $n = \log N$. 

\section{Excessive Scaling} \label{exscal}

\bt ($k$-NAE-SAT QAOA random assignment scaling exponent)
\begin{equation}
C^{\text{NAE}}_{p, k} (\underline{\beta}, \underline{0}) = C^{\text{NAE}}_{p, k} (\underline{0}, \underline{\gamma}) = 2^{1-k}r
\end{equation}
\begin{proof}
    First, we realise $\forall p$
    \begin{equation}
    \begin{cases}
        \ket{\Psi^{\text{NAE}}(\u{\beta}, \u{0}, \us)}_p = \prod\limits_{i = 0}^{p-1}\unb(\beta_i)\ket{+}^{\otimes n} &\propto \ket{+}^{\otimes n} = \frac{1}{\sqrt{2^n}} \sumb \ketx\\
        \ket{\Psi^{\text{NAE}}(\u{0}, \u{\gamma}, \us)}_p = \prod\limits_{i = 0}^{p-1}\uncsn(\gamma_i)\ket{+}^{\otimes n} &\propto \ket{+}^{\otimes n}= \frac{1}{\sqrt{2^n}} \sumb \ketx
    \end{cases}
    \end{equation}
    since in both cases, the resulting operators all commute and the state is simply effected with a global phase. As such, 
    \begin{equation}
    \begin{split}
        p^{\text{NAE}}_{succ}(\us) &= \frac{1}{2^n}\sum_{\ux, \ux^\prime \in \bn^n} \bra{\ux^\prime}\{ \hamcsn = 0 \}\ketx\\
        &= \frac{1}{2^n}\sum_{\ux, \ux^\prime \in \bn^n}\Pr(\ux \vdash_{\text{NAE}} \us) \delta_{\ux, \ux^\prime}\\ 
        &= \frac{1}{2^n}\sumb \Pr(\ux \vdash_{\text{NAE}} \us)
    \end{split}
    \end{equation}
    Since $\{ \hamcsn = 0 \}$ denotes the orthogonal projector onto the space of satisfying assignments.
    
    As $\us$ is in CNF
    \begin{equation}
        \Pr(\ux \vdash_{\text{NAE}} \us) = \prod_{i = 0}^{m-1} \Pr(\ux \vdash_{\text{NAE}} \sigma_i) = \prod_{i = 0}^{m-1}\left[1 - \Pr(\ux \nvdash_{\text{NAE}} \sigma_i)\right] 
    \end{equation}
    Recalling \ref{naesatassat}:
    \begin{equation}
        \Pr(\ux \nvdash_{\text{NAE}} \sigma_i) = \Pr\left(\ux \nvdash \sigma_i \vee \ux \vdash \bigwedge_{j = 0}^{k-1} l_{\sigma_{ij}}\right) = \left(\frac{1}{2}\right)^k + \left(\frac{1}{2}\right)^k = 2^{1-k}
    \end{equation}
    since all literals are chosen independently and satisfying a literal occurs with probability $1/2$ as it is negated with equal probability.

    As such,
    \begin{equation}
        p^{\text{NAE}}_{succ}(\us) = \frac{1}{2^n}\sumb \prod_{i = 0}^{m-1}\left[1 - \Pr(\ux \nvdash_{\text{NAE}} \sigma_i)\right]  = \frac{1}{2^n}\sumb \left[1 - 2^{1-k}\right]^{m(\us)} 
    \end{equation}
    since all the clauses are chosen independently. We write $m(\us)$ to emphasise $m$ is a property of the random instance $\us$.
    
    Finally, as $m \sim Poisson(rn)$
    \begin{equation}
    \begin{split}
        \mathbb{E}_{\us \sim CNF(n, k, r)}\left[ \left(1 - 2^{1-k} \right)^{m(\us)}\right] &= \mathbb{E}_{m \sim Poisson(nr)}\left[ \left(1 - 2^{1-k} \right)^{m}\right]\\
        &= \sum_{m \geq 0}\frac{e^{-rn}{(rn)}^m}{m!}(1 - 2^{1-k})^m\\
        &= e^{-rn}\sum_{m \geq 0}\frac{\left[rn(1 - 2^{1-k})\right]^m}{m!}\\
        &= e^{-rn}e^{rn(1 - 2^{1-k})}\\
        &= e^{-rn2^{1-k}}
    \end{split}
    \end{equation}
    This means that
    \begin{equation}
    \begin{split}
        \lim_{n \rightarrow \infty} \frac{1}{n} \log \mathbb{E}_{\us \sim CNF(n, k, r)}[p^{\text{NAE}}_{succ}(\us)] 
        = \lim_{n \rightarrow \infty} \frac{1}{n} \log e^{-rn2^{1-k}}
        = -2^{1-k}r
    \end{split}
    \end{equation}
    and so
    \begin{equation}
        C^{\text{NAE}}_{p, k} (\underline{\beta}, \underline{0}) = C^{\text{NAE}}_{p, k} (\underline{0}, \underline{\gamma}) = 2^{1-k}r
    \end{equation}
\end{proof}
\et 

\section{Additional Results} \label{fullresults}

\subsection{Success Probability}

Given parameters $\underline{\beta}^*$ and $\underline{\gamma}^*$, we evaluate the success probability of QAOA over $v = 2500$ \textbf{satisfiable} random $k$-NAE-SAT instances, $\{ \us_i : \us_i \sim CNF(n, k, \hat{r}^{\text{NAE}}_k)\}_{i = 0}^{v - 1}$ and calculate
\begin{equation}
    \hat{p}^{\text{NAE}}_{succ} = \frac{1}{v}\sum_{i = 0}^{v - 1} p^{\text{NAE}}_{succ}(\sigma_i) 
\end{equation}

\begin{figure}[h]
    \centering
    \includegraphics[width=\textwidth]{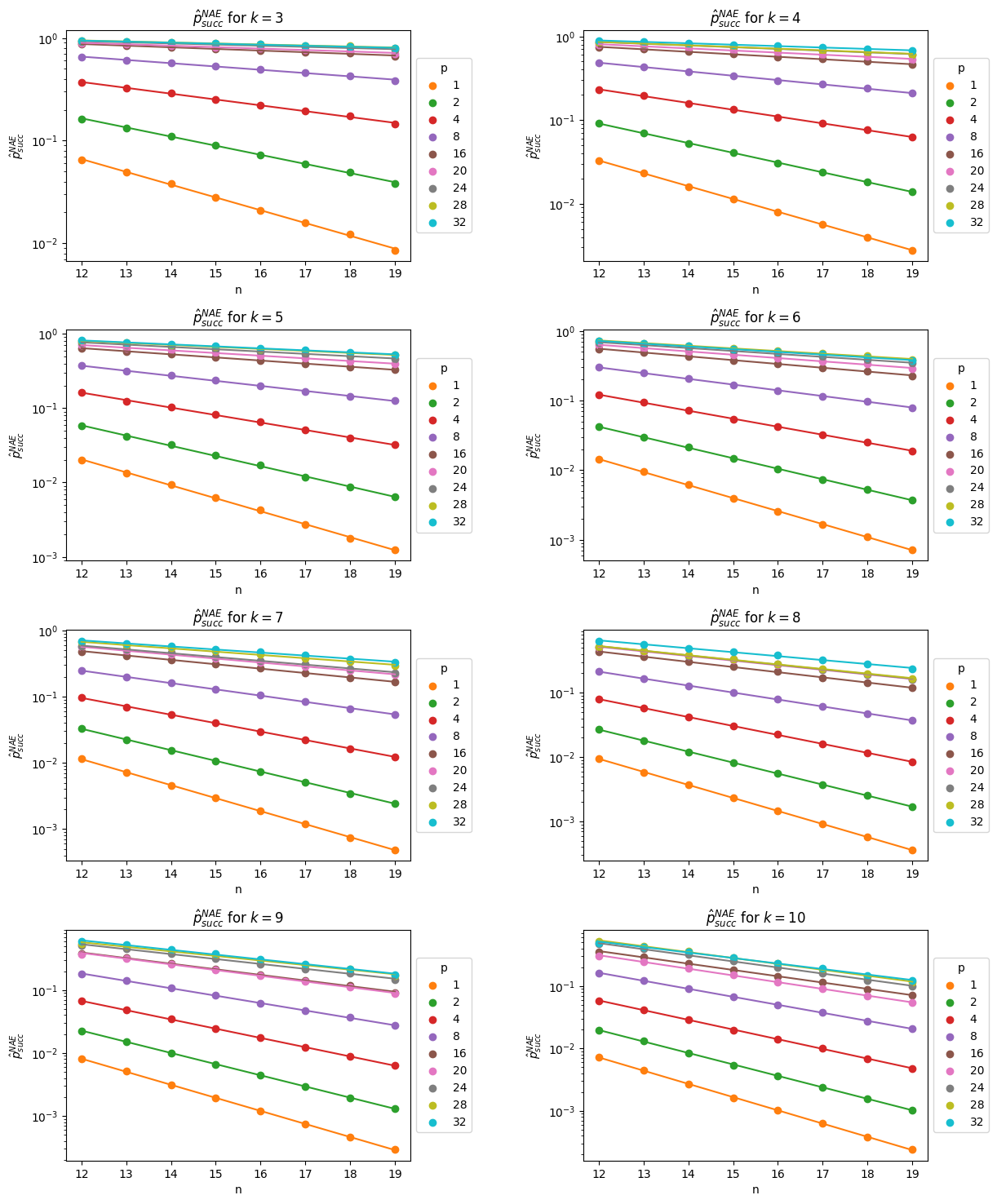}
    \caption{QAOA average success probabilities across 2500 satisfiable $k$-NAE-SAT $CNF(n, k, \hat{r}^{\text{NAE}}_{k})$ instances (error bars too small to be seen).}
    \label{fig:psucc_knaesat_allk}
\end{figure}

The results for $3 \leq k \leq 10$ and $p \in \{1,2,4,8,16,20,24,28,32\}$ are as in Figure \ref{fig:psucc_knaesat_allk}.

\subsection{Median Running Times}

Given parameters $\underline{\beta}^*$ and $\underline{\gamma}^*$, we evaluate the running time of QAOA over $v = 2500$ \textbf{satisfiable} random $k$-NAE-SAT instances, $r=\{ \us_i : \us_i \sim CNF(n, k, \hat{r}^{\text{NAE}})\}_{i = 0}^{v - 1}$, and calculate the median
\begin{equation}
    Med_{\us \in r}[r^{\text{NAE}}_{\us}]
\end{equation}

\begin{figure}[h]
    \centering
    \includegraphics[width=\textwidth]{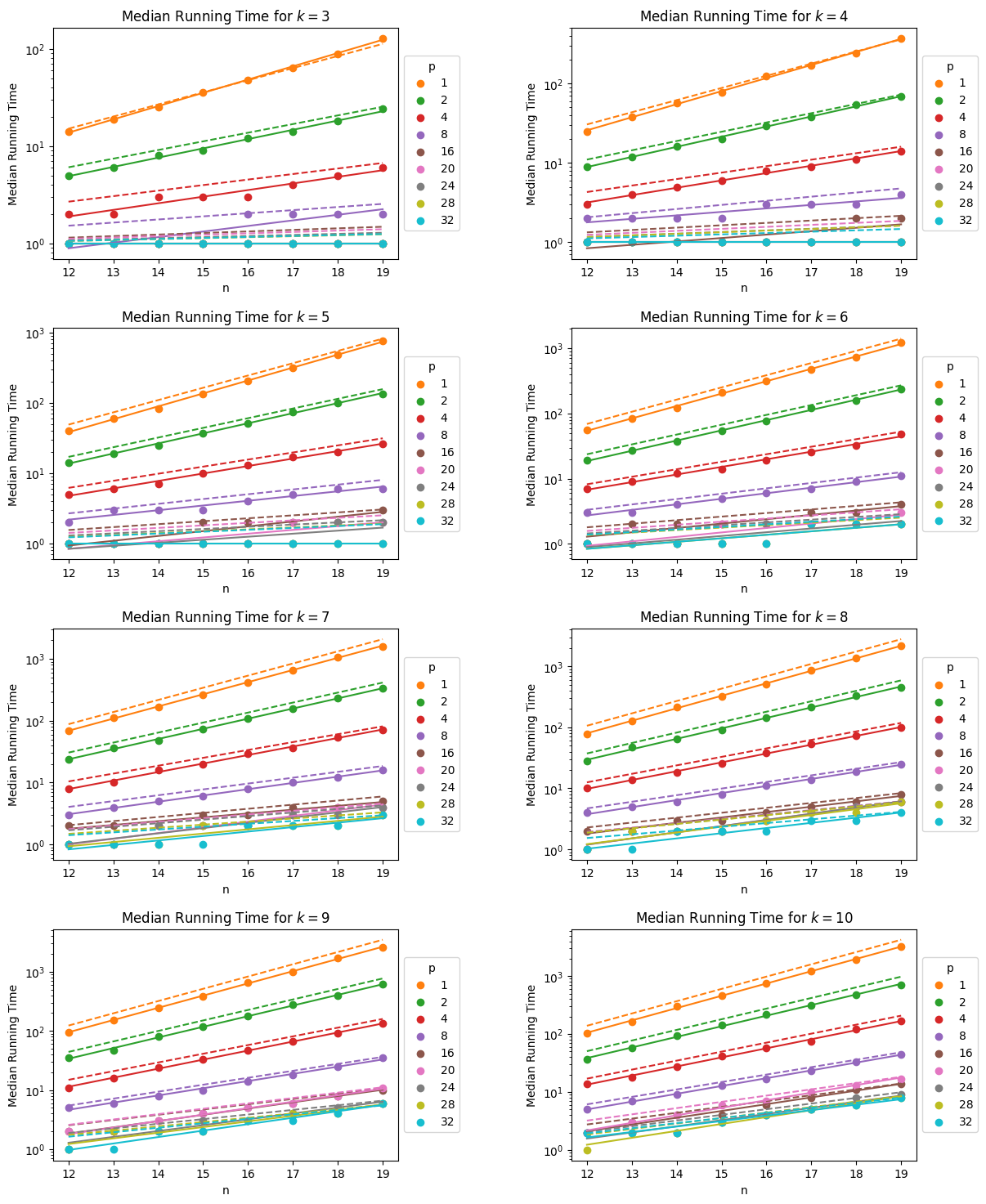}
    \caption{QAOA median running times across 2500 satisfiable $k$-NAE-SAT $CNF(n, k, \hat{r}^{\text{NAE}}_{k})$ instances. Dashed lines are the reciprocal of corresponding success probabilities.}
    \label{fig:mrt_knaesat_allk}
\end{figure}

The results for $3 \leq k \leq 10$ and $p \in \{1,2,4,8,16,20,24,28,32\}$ are as in Figure \ref{fig:mrt_knaesat_allk} where we compare the median running time to the reciprocal of the sucess probability.

\subsection{Benchmarking}

We compare the median running time of QAOA to that of \verb|WalkSATlm| and \verb|WalkSATm2b2| across 2500 \textbf{satisfiable} random $k$-NAE-SAT instances, $\{ \us_i : \us_i \sim CNF(n, k, \hat{r}^{\text{NAE}}_k\}_{i = 0}^{v - 1}$.

\begin{figure}[h]
    \centering
    \includegraphics[width=\textwidth]{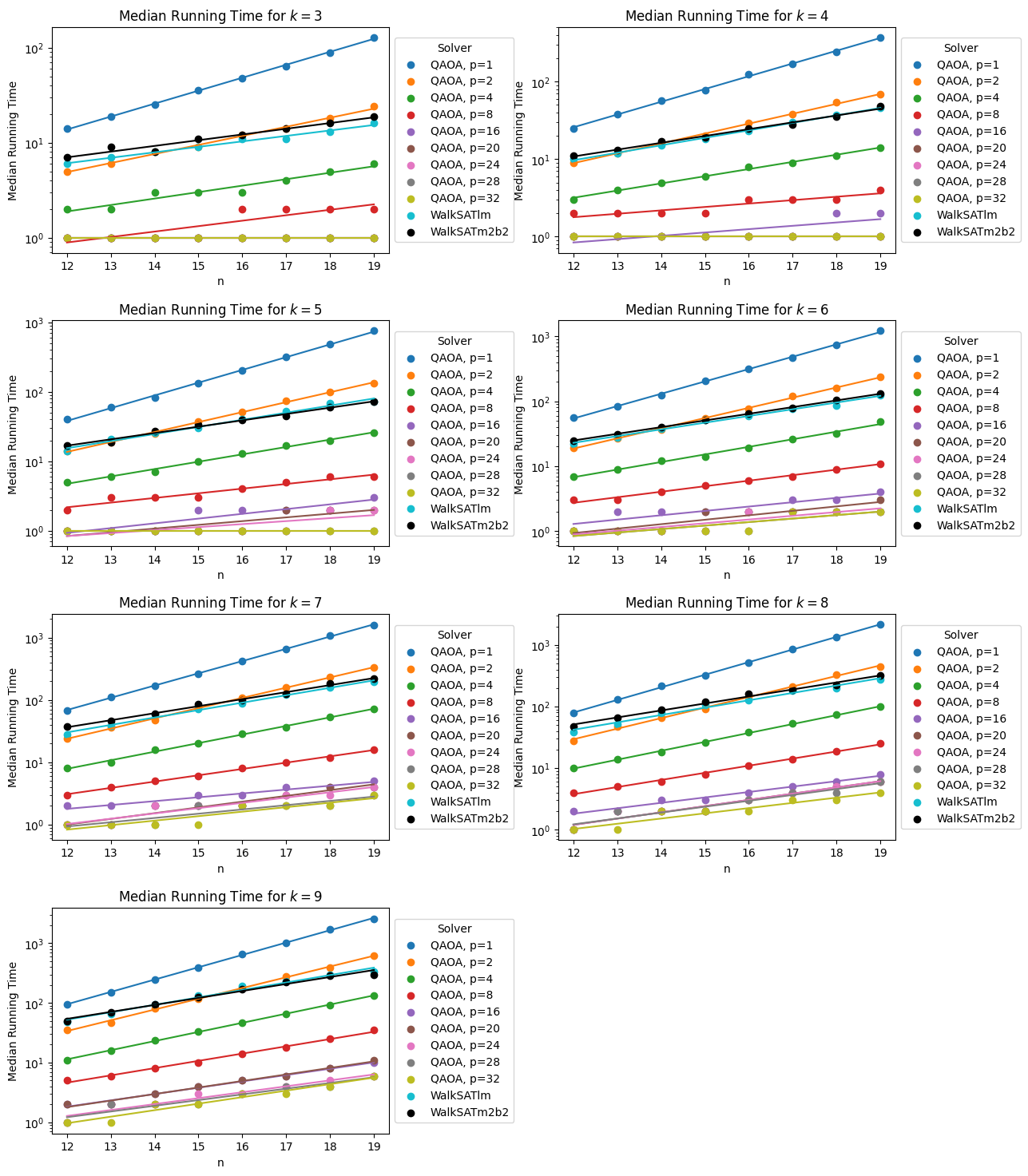}
    \cprotect\caption{Median running times across 2500 satisfiable $k$-NAE-SAT $CNF(n, k, \hat{r}^{\text{NAE}}_k)$ instances of QAOA, \verb|WalkSATlm| and \verb|WalkSATm2b2|.}
    \label{fig:benchmark_knaesat_allk}
\end{figure}

The results for $3 \leq k \leq 9$ and $p \in \{1,2,4,8,16,20,24,28,32\}$ are as in Figure \ref{fig:benchmark_knaesat_allk}. \footnote{Due to time and resource constraints we did not benchmark for $k=10$.}

\subsection{Excessive Scaling}

We assess the scaling $\Tilde{C}^{\text{NAE}}_{p, k} (\underline{\beta}, \underline{\gamma})$ of the above runtimes and fit the points to a power law $\sim ap^b$. This is compared to random assignment scaling $\Tilde{C}^{\text{NAE}}_{p, k} (\underline{\beta}, \underline{\gamma}) = 2^{1-k}r$.

\begin{figure}[h]
    \centering
    \includegraphics[width=\textwidth]{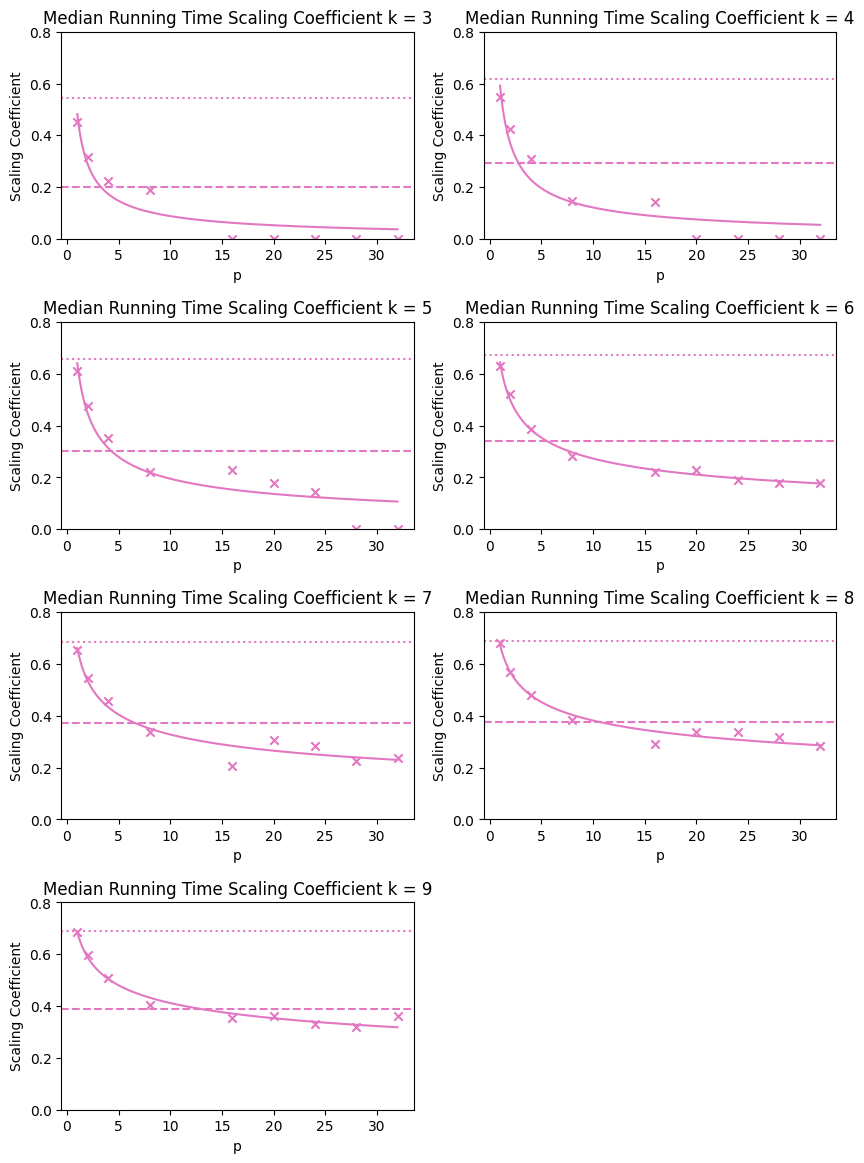}
    \cprotect\caption{Induced scaling of QAOA $k$-NAE-SAT median running times. Dashed line is observed \verb|WalkSATm2b2| scaling. Dotted line is random assignment scaling.}
    \label{fig:knaesat-excessive-scaling-allk}
\end{figure}

The results for $3 \leq k \leq 9$ and $p \in \{1,2,4,8,16,20,24,28,32\}$ are as in Figure \ref{fig:knaesat-excessive-scaling-allk} and Table \ref{powerlawfits}.

\begin{table}[h!]
\centering
\begin{tabular}{|c|c|c|}
\hline
$k$ & $a$ & $b$ \\
\hline
3 & 0.48193188 & -0.73653218 \\
4 & 0.59220998 & -0.68702534 \\
5 & 0.64106651 & -0.51678505 \\
6 & 0.64427816 & -0.37302963 \\
7 & 0.66099965 & -0.30455395 \\
8 & 0.67472907 & -0.24718223 \\
9 & 0.68581328 & -0.22158058 \\
10 & 0.6987176 & -0.19971659 \\
\hline
\end{tabular}
\cprotect\caption{Power law fit $ \Tilde{C}^{\text{NAE}}_{p, k} (\underline{\beta}, \underline{\gamma}) \sim ap^b$\label{powerlawfits}.}
\end{table}

\section{Software}
We identify that general satisfiability problems, including SAT and its variants, fit into a \textit{clause-literal} framework. The instances consist of clauses that have to be satisfied in some combination, e.g. all of them in $k$-SAT (since the formula is in CNF). Similarly, the clauses consist of literals that have to be satisfied in some combination, e.g. any of them in $k$-SAT or at least one, but not all, in $k$-NAE-SAT. As such, we make use of an object oriented approach to encode our problem instances, wherein the responsibility of confirming satisfiability is delegated to the problem instance.

In particular, all problems derive from a base \verb|Formula| class that contains a set of base \verb|Clause| instances. The \verb|Formula| class implements the \verb|is_satisfied| method, accepting a bitstring and returning \verb|True| or \verb|False| to denote whether the instance is satisfied by the assignment the bitstring represents. We create extensions of this base class to represent the requirements of the formula, such as \verb|CNF| which encodes that the formula is satisfied if and only if all the clauses are satisfied. Similarly, extensions of the \verb|Clause| class: \verb|DisjunctiveClause| and \verb|NAEClause| represent each clauses' constraints.

Not only does this allow us to share logic between classes, but it also means we can construct algorithms that are agnostic to problem instance specifics until run time. In addition, we can generate random problem instances using a common selection procedure that simply instantiates the relevant problem class. We validate the correctness of all this, including our algorithms and problem representations, using the Python \verb|unittest| framework.

The simulations are implemented in \verb|Qiskit| \cite{Qiskit}, for the generalised unitary, and in \verb|PyTorch| \cite{PyTorch}, for the diagonalised unitary. In addition, using \verb|Condor| \cite{Condor} on Imperial College's Department of Computing systems, we are able to parallelise up to 300 simultaneous processes, each with 16 threads.

\end{document}